 \documentclass[authoryear,preprint,review,12pt]{elsarticle}

\usepackage{amssymb}
\usepackage{amsthm}
\usepackage{amsmath}
\usepackage{lineno}
\usepackage{soul}
\usepackage{subfig}
\usepackage{algorithm}
\usepackage{algpseudocode}

\usepackage{color}

\newcommand{\comment}[1]{#1}

\journal{Computational Statistics \& Data Analysis}

\begin{document}

\begin{frontmatter}



\title{An Induced Natural Selection Heuristic for Finding Optimal Bayesian Experimental Designs}


\author[cam,uofm,doherty]{David J. Price\corref{cor1}}
\cortext[cor1]{david.j.price@alumni.adelaide.edu.au}

\author[uofa,acems]{Nigel G. Bean}

\author[uofa,acems]{Joshua V. Ross}

\author[uofa,acems]{Jonathan Tuke}

\address[cam]{Disease Dynamics Unit, Department of Veterinary Medicine, University of Cambridge, Madingley Road Cambridge CB3 0ES, United Kingdom}
\address[uofm]{Centre for Epidemiology and Biostatistics, Melbourne School of Population and Global Health, The University of Melbourne, VIC 3010, Australia}
\address[doherty]{Victorian Infectious Diseases Reference Laboratory Epidemiology Unit at the Peter Doherty Institute for Infection and Immunity, The University of Melbourne and Royal Melbourne Hospital, VIC 3000, Australia}
\address[uofa]{School of Mathematical Sciences, University of Adelaide, SA 5005, Australia}
\address[acems]{ARC Centre of Excellence for Mathematical \& Statistical Frontiers, School of Mathematical Sciences, University of Adelaide, SA 5005, Australia}

\begin{abstract}
Bayesian optimal experimental design has immense potential to inform the collection of data so as to subsequently enhance our understanding of a variety of processes. However, a major impediment is the difficulty in evaluating optimal designs for problems with large, or high-dimensional, design spaces. We propose an efficient search heuristic suitable for general optimisation problems, with a particular focus on optimal Bayesian experimental design problems. The heuristic evaluates the objective (utility) function at an initial, randomly generated set of input values. At each generation of the algorithm, input values are ``accepted'' if their corresponding objective (utility) function satisfies some acceptance criteria, and new inputs are sampled about these accepted points. We demonstrate the new algorithm by evaluating the optimal Bayesian experimental designs for the previously considered death, pharmacokinetic and logistic regression models. Comparisons to the current ``gold-standard'' method are given to demonstrate the proposed algorithm as a computationally-efficient alternative for moderately-large design problems (i.e., up to approximately 40-dimensions).
\end{abstract}


\end{frontmatter}

\section{Introduction}
\label{section:intro}

Optimising the design of an experiment is an important consideration in many areas of science, including, but not limited, to: biology \citep{Faller:2003}, clinical trials \citep{Berry:2004} and epidemiology \citep{Pagendam:2013}.  The theory of optimal experimental design is a statistical framework that allows us to determine the optimal experimental protocol to gain the most information about model parameters, given constraints on resources. 

In evaluating an optimal Bayesian design, there are two main components: the search across the design space, and the evaluation of the utility. There have been many approaches to improving the efficiency of both aspects, summarised by \citet{Ryan:2015}. Recently, \citet{Overstall:2017} proposed the Approximate Coordinate Exchange (ACE) algorithm to address the search aspect of the Bayesian experimental design problem. The method utilises a coordinate exchange algorithm to update one dimension of the design at a time, coupled with a Gaussian process in order to search each dimension efficiently. It has been asserted that the future of optimal Bayesian experimental design lies in the ability to evaluate the optimal designs for large-scale problems (\emph{i.e.}, large or high-dimensional design spaces), in a computationally-efficient manner \citep{Ryan:2015}. In this paper, we address this by proposing a new search algorithm targeted at finding optimal Bayesian experimental designs.

The search heuristic we present performs targeted sampling of the design space to find high utility designs, without making any assumptions about the shape of the utility function. An initial population of random designs is generated -- synonymous with multiple algorithm runs from random initial conditions as in other optimisation routines. Our method borrows the idea of targeting regions of high utility, as per the MCMC approach of \citet{Muller:1999}, by sampling new designs at each iteration around the ``best'' designs; chosen according to some acceptance criteria.
We describe this algorithm using the notion of ``survival-of-the-fittest", as the ``fittest'' individuals -- according to their objective (utility) function value -- survive at each iteration (generation) based on a user-defined acceptance criteria, to produce offspring for the next generation. Hence, we propose this as a new type of evolutionary algorithm (\emph{e.g.}, \citealp{Goldberg:1989}), and refer to it herein as the Induced Natural Selection Heuristic (INSH).

By independently sampling new designs around each accepted design, we aim to avoid the pitfalls associated with some other optimisation routines. For example, INSH is able to sample multiple regions of high utility at a time, thus exploring multiple local optima simultaneously, rather than potentially being stuck at a single local optima. Furthermore, by not combining the retained designs in any way, INSH avoids the potential to move to a region of low utility that is at the ``centre" of multiple local optima -- as may occur in a cross-entropy or genetic algorithm. By taking a sampling approach, as opposed to trying to approximate the function, INSH makes no assumptions about the shape of the utility function -- thus, it is not limited to utility functions that are, for example, smooth. Utilising (embarrassingly) parallel computation tools, the method can efficiently evaluate the utility for a large number of designs in each iteration.

The ACE algorithm has allowed the consideration of Bayesian optimal designs for a larger, more-complex class of statistical models and experiments than was possible with previous algorithms. There are a number of drawbacks to ACE, however. By searching in one-dimension at a time, ACE risks missing the globally-optimal design, and instead may find only local optima. An approach to avoid this is to re-run the algorithm from a number of randomly generated initial designs \citep{Overstall:2017}. Similarly, as noted by the authors, by searching in one-dimension at a time, the algorithm will be inefficient in scenarios where there is a large correlation between the design variables -- a problem which adds to the difficulty in choosing a suitable number of iterations for each phase of the algorithm. The algorithm requires a sufficiently-good estimate of the utility when determining whether to accept the candidate design -- spurious estimates may lead to sub-optimal candidate designs being accepted, and thus push the algorithm away from regions of high utility. Alternatively, a large improvement in the computation time arises from the estimation of the utility surface in each dimension in the form of a Gaussian process based on a number of candidate points. This approximation to the utility surface based on noisy evaluations of the utility aims to provide a smooth approximation to the surface. When the surface is not smooth, or has a discontinuity (\emph{e.g.}, as exists in the utility surface for the death model in Figure \ref{deathmodel:fullutilitysurface} at $\boldsymbol{t}\approx(2.75,t_2)$ and $\boldsymbol{t}\approx(t_1,2.75)$), this has the potential to cause problems for the ACE algorithm.
 
In the following, we present the INSH search algorithm in a general framework, and we note that efficient evaluation of the utility is another problem that needs to be addressed. We consider two existing approaches to evaluating the utility: an Approximate Bayesian Computation (ABC) approach used by \citet{Price:2016}, in a scenario where the benefits of this approach are realised; and a nested Monte-Carlo approximation using code from the \verb+acebayes+ package \citep{Rpack:acebayes}, otherwise.

We consider the problem of finding the optimal design for the death model, a pharmacokinetic (PK) model tracking the concentration of a drug or treatment in the blood, and a four-factor logistic regression model. In the death and PK examples, a design $d$ consists of $n$ sampling times $(t_1,\dots,t_n)$, subject to some problem-specific constraints. First, we address the question of when to observe the stochastic process in order to gain the most information about the model parameters governing the death model. The Markovian death model has been considered previously in a Bayesian framework by \citet{Cook:2008}, \citet{Drovandi:2013}, and \citet{Price:2016}. We compare the optimal designs for 1-4 observation times in order to demonstrate the efficacy of the method. Second, we consider the question of sampling times for a PK model -- a process where the design space is higher-dimensional -- in order to demonstrate the efficiency of the INSH algorithm for larger design spaces. The optimal designs are compared to those evaluated using the ``gold-standard'' Approximate Coordinate Exchange (ACE) algorithm of \citet{Overstall:2017}. We also consider the idea of sampling windows for this example, which have been considered previously by \citet{Duffull:2003}, \citet{Chenel:2005}, \citet{Graham:2006}, \citet{McGree:2012}, and \citet{Duffull:2012}, for example. 
Finally, we compare the results of the INSH algorithm to those of the ACE algorithm for a standard four-factor logistic regression model \citep{Overstall:2017} -- a considerably higher-dimensional problem. We consider examples with $n=6,10,24$, and $48$ (independent) replicates in each experiment; corresponding to a design space with up to 192 dimensions (i.e., when $n=48$ replicates). 

\subsection{Bayesian Optimal Experimental Design}

The aim of optimal experimental design is to determine the best experimental protocol in order to maximise some utility of the experiment. To achieve this aim, we specify a utility function $U(\boldsymbol{\theta},\boldsymbol{y},d)$ representing how we `value' the experimental design $d$, chosen from the set of all designs $\mathcal{D}$, where $\boldsymbol{\theta}$ represents the model parameters and $\boldsymbol{y}$ is the data. We are interested in the expected utility of using design $d$, over the unknown model parameters and data. That is, we wish to evaluate,
\begin{align}
	u(d) &= E_{\boldsymbol{\theta},\boldsymbol{y}}[ U(\boldsymbol{\theta},\boldsymbol{y}, d)] \notag \\
		&=  \int_{\boldsymbol{y}} \int_{\boldsymbol{\theta}} U(\boldsymbol{\theta},\boldsymbol{y}, d) p(\boldsymbol{y} \mid \boldsymbol{\theta},d) p(\boldsymbol{\theta}) d\boldsymbol{\theta} d\boldsymbol{y},
	\label{utilitydefn}
\end{align}
where $p(\boldsymbol{y} \mid \boldsymbol{\theta}, d)$ is the likelihood function of the unobserved data $\boldsymbol{y}$, under design $d$, and $p(\boldsymbol{\theta})$ is the prior distribution of the model parameters. The optimal design $d^*$ maximises the expected utility over the design space $\mathcal{D}$, that is, $d^* = \text{argmax}_{d\in\mathcal{D}} u(d)$. The utility function we use throughout this work is the Kullback-Leibler divergence \citep{Kullback} from the prior distribution to the posterior distribution (which is independent of $\boldsymbol{\theta}$), 
\begin{equation*}
U(\boldsymbol{y}, d) = \int_{\boldsymbol{\theta}} \log \left( \frac{p(\boldsymbol{\theta} \mid \boldsymbol{y},d)}{ p(\boldsymbol{\theta})}  \right) p(\boldsymbol{\theta} \mid \boldsymbol{y}, d) d\boldsymbol{\theta},
\end{equation*}
which leads to an expected utility:
\begin{align}
	u(d)  = \int_{\boldsymbol{y}} \int_{\boldsymbol{\theta}} \log \left( \frac{p(\boldsymbol{\theta} \mid \boldsymbol{y},d)}{ p(\boldsymbol{\theta})}  \right) p( \boldsymbol{y} \mid  \boldsymbol{\theta},d) p(\boldsymbol{\theta})  d\boldsymbol{\theta} d\boldsymbol{y}. 
	\label{ekld}
\end{align}
See \citet{Price:2016} for details of the derivation.  Alternatively, it is commonplace to consider the Shannon Information Gain (SIG), which can be written as:
\begin{align}
U(\boldsymbol{\theta}, \boldsymbol{y}, d) =& \log p(\boldsymbol{\theta} \mid \boldsymbol{y}, d) - \log p(\boldsymbol{\theta}) \notag \\
								=& \log p(\boldsymbol{y}  \mid \boldsymbol{\theta}, d) - \log p(\boldsymbol{y} \mid d),
\label{eqn:SIG}
\end{align}
through the application of Bayes' theorem. Maximisation of the expected SIG is equivalent to maximisation of the expected Kullback-Leibler divergence above. Unfortunately, it is often not possible to obtain an analytic evaluation of the expected utility function $u(d)$ (Equation \eqref{utilitydefn}), and approximate methods are required (see Section \ref{subsection:evaluate_utility}). 

\subsection{ACE Algorithm}

The Approximate Coordinate Exchange algorithm of \citet{Overstall:2017} directly addresses the need for a computationally-efficient algorithm for determining optimal Bayesian experimental designs in high-dimensional design spaces \citep{Ryan:2015}. The reader is directed to \citet{Overstall:2017} for full details of the algorithm. Briefly, the algorithm considers each dimension of the experimental design one-at-a-time (\emph{e.g.}, the first observation time in an observation schedule), and evaluates the utility at a number of new, candidate values in that dimension (\emph{e.g.}, consider the utility at each of $q$ equally-spaced times across the feasible range of observation times, conditional on the other elements of the design). Having obtained these approximate utilities across the feasible range for the particular dimension of the design, a Gaussian process is fit to these candidate values to find an approximate ``optimal'' value as an update to this dimension of the design (accepted with some probability). The algorithm cycles through each design variable (probabilistically) updating them to the best value according to the Gaussian process approximation to the utility.  The ACE algorithm is the first algorithm capable of dealing with high-dimensional design problems, in a computationally feasible amount of time.

\section{Proposed Method: INSH Algorithm}

In the following, we present a new algorithm to find optimal Bayesian experimental designs efficiently. We describe an algorithm that can utilise the current advantages of parallel computing -- which are rapidly improving as parallel-computing becomes more widely-available, more easy to implement, and more powerful. Simultaneously, we embrace an advantageous aspect of the inherently sequential, and thus difficult to parallelise efficiently, MCMC algorithms implemented by \citet{Muller:1999}, \citet{Cook:2008}, and \citet{Drovandi:2013}: namely, we seek to spend less computational effort evaluating designs in low-utility regions. This forms the crux of the efficiency of an MCMC approach, and is achieved by sampling from a function proportional to the utility. The new algorithm we propose instead evaluates the utility of multiple designs simultaneously -- in order to realise the benefits of parallel computing -- and samples new designs at each iteration of the algorithm around designs that satisfy some acceptance criteria. The acceptance criteria for designs at each iteration can be chosen in a number of different ways. In this paper, we demonstrate the acceptance of a fixed number of the ``best'' designs, similar to the proportion of ``elite'' samples in a cross-entropy algorithm \citep{DeBoer:2005}. In contrast to these existing optimisation algorithms, the algorithm presented here considers multiple designs at each iteration, allowing us to explore the design space more efficiently. The INSH algorithm is detailed in Algorithm \ref{INSH_algorithm}. 
Note that in Step 6, the best design considered in any previous iteration is reintroduced into the set of designs that are to be sampled around, in order to continue to explore this region.
\begin{algorithm}[h]
\caption{INSH Algorithm}\label{INSH_algorithm}
\begin{algorithmic}[1]
	\State Choose an initial set of designs, $D$ (\emph{e.g.}, a coarse grid of design points across the design space, or randomly sample).
	\State Specify the number of generations (iterations) of the algorithm $W$, a perturbation function $f(d\mid d')$, and the acceptance criteria.
	\For{$w=1$ to $W$}
	\State \parbox[t]{0.95\textwidth}{\strut For each design $d^i\in D$, sample parameters $\boldsymbol{\theta}\sim p(\boldsymbol{\theta})$, and simulate data $\boldsymbol{y}^i$ from the model.\strut }
	\State \strut Evaluate utility $u(d^i)$, for each design $d^i \in D$.  \label{insh_evaluate_utility}
	\State \parbox[t]{0.95\textwidth}{\strut Set $D'$ to be the designs which satisfy the acceptance criteria, and the current optimal design $d^*$ (even if it occurred in a previous generation).\strut }
	\State \parbox[t]{0.95\textwidth}{\strut Sample $m$ designs from $f(d\mid d')$, for each $d'\in D'$. Set $D$ to be these newly sampled designs.\strut }
	\EndFor
	\Ensure Set of designs $d$, and corresponding approximate utilities $u(d)$ (and hence, the optimal design $d^* = \underset{d\in\mathcal{D}}{\text{argmax}}(u(d))$).
\end{algorithmic}
\end{algorithm}

\subsection{Evaluation of the Utility}
\label{subsection:evaluate_utility}
An efficient approach to evaluate the utility of a design in Step \ref{insh_evaluate_utility} of Algorithm \ref{INSH_algorithm} is that of the ABCdE algorithm \citep{Price:2016}. \comment{We state the ABCdE algorithm in Algorithm 2 of Online Resource A, and direct the reader to \citet{Price:2016} for a complete description of the algorithm}. In particular, we use Steps 3 to 9 of Algorithm 2, in Online Resource A. Note that this approach is suitable for discrete data, and for low-dimensional design spaces. This is due to the majority of the efficiency coming from having to evaluate a posterior distribution only once for each unique data set. As the number of possible unique data sets increases -- for example, either by observing the process more often (increasing the size of the design space), or having a larger population -- this approach to evaluating the utility becomes less efficient. We use this approach to evaluate the utility for the Markovian death model, in order to demonstrate the INSH algorithm.

For cases where the dimension of the data is too large (or continuous), we must consider an alternative approach to evaluating the utility for each design. As noted previously, this is one of the two main challenges when searching for optimal Bayesian designs. A suitable and efficient method for evaluation of the utility for a design is often problem-specific, and a number of different approaches have been considered -- a summary of these approaches can be found in \citet{Ryan:2015}. For the PK and logistic regression examples we consider subsequently, we implement the utility function of \citet{Overstall:2017}, as provided in the \verb+acebayes+ package in R \citep{Rpack:acebayes}. Briefly, the SIG utility in equation \eqref{eqn:SIG}, is estimated by a nested Monte-Carlo approximation of the values $p(\boldsymbol{y}  \mid \boldsymbol{\theta}, d)$ and $p(\boldsymbol{y} \mid d)$, within the Monte-Carlo approximation to the expected utility, ${u}(d)$. Borrowing the notation of \citet{Overstall:2017}, define $\boldsymbol{\psi} = (\boldsymbol{\theta}, \boldsymbol{\gamma})$ to be the combination of the parameters of interest, $\boldsymbol{\theta}$, and nuisance parameters, $\boldsymbol{\gamma}$. Then, we use $\tilde{B}$ simulations to approximate the inner Monte-Carlo estimates:
\begin{align*}
\tilde{p}(\boldsymbol{y} \mid \boldsymbol{\theta}, d) = \frac{1}{\tilde{B}} \sum_{b=1}^{\tilde{B}} p(\boldsymbol{y} \mid \boldsymbol{\theta}, \tilde{\boldsymbol{\gamma}}_b, d ),  \quad \text{and } \quad
\tilde{p}(\boldsymbol{y} \mid d) = \frac{1}{\tilde{B}} \sum_{b=1}^{\tilde{B}} p(\boldsymbol{y} \mid \tilde{\boldsymbol{\theta}}_b, \tilde{\boldsymbol{\gamma}}_b,d ),
\end{align*}
where $(\tilde{\boldsymbol{\theta}}_b, \tilde{\boldsymbol{\gamma}}_b)$ are the $\tilde{B}$ parameters sampled from the prior distribution of $\boldsymbol{\psi}$. Similarly, $B$ simulations are used to evaluate the outer Monte-Carlo estimate,
\begin{align*}
\tilde{u}(d)=\frac{1}{B} \sum_{l=1}^B\left[  \log \tilde{p}(\boldsymbol{y}_l  \mid \boldsymbol{\theta}_l, d) - \log \tilde{p}(\boldsymbol{y}_l \mid d) \right],
\end{align*}
with $\{\boldsymbol{y}_l, \boldsymbol{\theta}_l\}$ parameters, and corresponding simulations, sampled from the prior and simulated from the model, respectively. 
In the work of \citet{Overstall:2017}, the authors use $\tilde{B}=B=1{,}000$ to evaluate the candidate designs' utilities in the one-dimensional search (Step 1b of the ACE Algorithm in \citealp{Overstall:2017}), and $\tilde{B}=B=20{,}000$ to evaluate the utility when determining whether to accept the candidate design (Steps 1d and 3e of the ACE Algorithm in \citet{Overstall:2017}; note that Step 3 is not implemented for the compartmental model). 

\subsection{Choice of Acceptance Criteria}

There are a number of ways in which we can choose to retain designs; taking inspiration from other optimisation routines. For example, one could retain all designs that are greater than some percentage of the current maximum (\emph{e.g.}, retain all designs that have at least $95\%$ of the information compared to the current ``optimal"), although this approach requires some insight to how ``flat" the utility surface is in order to avoid retaining too many or too few designs at each iteration. While we do not present the results here, testing this approach for the Markovian death model showed promising results.

The approach that we implement in this work is similar to the ``elite" samples of a cross-entropy algorithm \citep{DeBoer:2005}. That is, at each generation, the algorithm accepts the best $r$ designs according to their utility. At the next generation of the algorithm, we sample $m$ designs from the perturbation kernel from each of these $r$ designs. In order to balance the trade-off between exploration and exploitation, one can specify a sequence of decreasing and increasing values for $m_w$ and $r_w$, respectively. Specifying the number of designs that are retained and sampled at each iteration ensures full control over the number of designs considered at each generation of the algorithm, allowing specification of the computational effort spent in searching for the optimal design. Thus, one may reasonably evaluate the optimal (or near-optimal) Bayesian design in a computationally-efficient time-frame. 

For the high-dimensional design spaces considered in the logistic regression example, we choose to modify this acceptance step slightly. Specifically, we choose to retain the best $r_w$ designs from the current \emph{and} previous iterations of the INSH algorithm. This acts as a failsafe, in instances where the newly proposed designs end up in regions of lower utility than the original (retained) design, which is more likely to occur with particular high-dimensional designs. This acceptance criteria allows the algorithm to start again from the previous iteration (for a subset of the designs), rather than end up missing regions of high utility through a poor round of sampling.

\subsection{Perturbation Kernel}

The \emph{perturbation kernel} is a probability distribution used to sample new designs at each generation of the INSH algorithm, by \emph{perturbing} (\emph{i.e.}, adding some noise to) previous designs. In the death and PK examples we consider in this work, we use a truncated, multivariate-Normal distribution (where the dimension is given by the dimension of the design space, and the truncation is to ensure constraints are satisfied). For the logistic regression example, we demonstrate the flexibility of choice in this aspect, by using a uniform distribution centred on each design point. \comment{There are no explicit guidelines on how to choose the kernel, and the choice is often driven by knowledge and experience of the problem at hand, as with sequential importance sampling methods (e.g., \citealp{Toni:2009}), however one can reasonably sample from any distribution, centred on the current design points, which can suitably explore the design space. The authors propose that without any knowledge of the relationship between the design points, a symmetric perturbation kernel is a sensible starting point.} A standard cross-entropy algorithm uses the accepted samples to define the mean and (co-) variance structure of a (multivariate-) Normal distribution, and all new samples are generated by this distribution. We prefer to avoid this approach, instead allowing the region surrounding each accepted point to be explored individually. Combining all accepted samples into a single distribution from which to sample, may result in new samples not being generated in regions of high utility (for example, when considering multi-modal utility surfaces), and requires re-evaluation of the (co-)variance matrix at each generation.

\subsection{Stopping Criteria}
A common feature of optimisation tools is a criterion for stopping the algorithm. It would be straight-forward for the user to implement a stopping criteria based on the change in utility of newly sampled designs at each iteration of the algorithm, based on the level of accuracy desired. In the examples considered herein, we choose to demonstrate the algorithm by running it for a fixed number of iterations, and assessing convergence graphically through box-plots of the estimated utility across each generation of the algorithm (similar to the trace plots of \citealp{Overstall:2017}).  

\section{Examples}
\subsection{Markovian Death Model}
Consider the Markovian death model as defined by \citet{Cook:2008}. There is a population of $N$ individuals which, independently, move to an infectious class $I$ at constant rate $b_1$ -- for example, due to infection from an environmental source. The Markov chain models the number of infectious individuals at time $t$, $I(t)$ (where the number of susceptible individuals is $S(t)=N-I(t)$). The positive transition rates of the Markov chain are given by $q_{i,i+1}=b_1(N-i)$, for $i=0,\dots,N-1$. The prior distribution we consider is $b_1 \sim \log\text{-}N(-0.005, 0.01)$, chosen such that the mean lifetime of individuals in the population is one, with an approximate variance of 0.01 (as per \citealp{Cook:2008}).

The optimal experimental design for the Markovian Death model has previously been considered in a Bayesian framework by \citet{Cook:2008}, \citet{Drovandi:2013}, and \citet{Price:2016}. \citet{Cook:2008} utilised the MCMC approach of \citet{Muller:1999}, and used an exact posterior, hence, the designs of \citet{Cook:2008} provide a gold-standard with which to compare our results. \citet{Drovandi:2013} also utilised the MCMC approach of \citet{Muller:1999}, however, coupled with an approximate posterior distribution evaluated via an ABC approach. We note however, that the MCMC approach struggles to evaluate the optimal design once considering more than four design parameters. This is due to the increasing computational difficulty associated with the evaluation of the mode of the multi-dimensional utility surface (\citealp{Drovandi:2013}). \citet{Price:2016} provided an exhaustive-search across a grid on the design space, where the utility was evaluated using the ABCdE method. The INSH code for the death model is implemented in MATLAB R2015b. 

\subsection{Pharmacokinetic Model}
Consider the PK experiment considered by \citet{Ryan:2014} and \citet{Overstall:2017}. In these PK experiments, individuals are administered a fixed amount of a drug. Blood samples are taken in order to understand the concentration of the drug within the body over time. 

Let $y_t$ represent the observed concentration of the drug at time $t$. We model the concentration as $ y_t = \mu(t)(1 + \epsilon_{1t}) + \epsilon_{2t}$, where, 
$$\mu(t) = \frac{400 \theta_2}{\theta_3 (\theta_2 - \theta_1)} \left( e^{-\theta_1 t} - e^{-\theta_2t} \right),$$ 
is the mean concentration at time $t$, and $\epsilon_{1t} \sim N(0, \sigma^2_{prop})$, $\epsilon_{2t} \sim N(0, \sigma^2_{add})$, $\sigma^2_{prop}=0.01$ and $\sigma^2_{add}=0.1$. That is, $y_t \sim N\left( \mu(t),  \sigma_{add}^2 + \sigma_{prop}^2\mu(t)^2 \right).$

The blood samples are taken within the first 24 hours after the drug is administered (that is, $t\in[0,24]$), and it is not practical to take blood samples less than 15 minutes apart (hence, $t_{i+1}-t_i\geq0.25$). We wish to make 15 observations of this system in order to obtain information about the model parameters $\boldsymbol{\theta} = (\theta_1, \theta_2, \theta_3)$, where $\theta_1$ represents the first-order elimination rate constant, $\theta_2$ represents the first-order absorption rate constant, and $\theta_3$ represents the \emph{volume of distribution} -- a theoretical volume that a drug would have to occupy in order to provide the same concentration as is currently present in the blood plasma, assuming the drug is uniformly distributed \citep{Ryan:2014}.

As per \citet{Ryan:2014} and \citet{Overstall:2017}, the model parameters $\boldsymbol{\theta}=(\theta_1,\theta_2,\theta_3)$ are assumed \emph{a priori} to be independently, normally distributed on the log-scale, with mean $\log(0.1)$, $\log(1)$, and $\log(20)$ respectively, and variance 0.05. \citet{Duffull:2012}, \citet{McGree:2012},  \citet{Ryan:2014}, and \citet{Ryan:2015pk} have previously evaluated optimal Bayesian experimental designs for pharmacokinetic models, either for only a few sampling times ($<5$), or more sampling times via dimension reduction schemes (\emph{e.g.}, search across the two-parameters of a Beta distribution, where the quantiles are scaled to give the observation times). \citet{Overstall:2017} are currently the only example of a method efficient enough to establish optimal Bayesian designs for a design-problem of this magnitude directly (i.e., without implementing a dimension reduction scheme), in a feasible amount of computation time.

Furthermore, we show how the output of the INSH algorithm can be used simply to construct \emph{sampling windows} -- a range of values for each observation, rather than a fixed value for each observation time. 
Sampling windows allows those implementing an optimally-chosen design some flexibility in choosing the sampling times, such that the resulting design is more practically feasible. By defining sampling windows, we can dictate a set of near-optimal designs -- which are practically feasible -- which can be implemented more easily. This avoids the scenario where an inferior design is chosen preferentially by those that are implementing the design, having been supplied with an impractical optimal design. 
Sampling windows have been considered previously for similar types of models, for example, in \citet{Duffull:2003}, \citet{Chenel:2005}, \citet{Graham:2006}, \citet{Duffull:2012}, and \citet{McGree:2012}, to name a few. As the output of the INSH algorithm consists of a large number of designs sampled around regions of high utility -- as opposed to a single design, as in ACE -- the construction of sampling windows is a simple extension to the algorithm. The INSH algorithm for the PK example is implemented in R (version 3.3.0).

\subsection{Logistic Regression in Four Factors}

Finally, we consider the logistic regression model of \citet{Overstall:2017} in order to demonstrate the benefits of INSH for a considerably higher-dimensional design problem. We consider only the case with independent groups (\emph{i.e.}, no random effects). Let $y_s \sim \text{Bernoulli}(\rho_s)$ be the $s^{th}$ response ($s=1,\dots,n$), and,
\begin{equation*}
\log \left(  \frac{\rho_{s}}{1-\rho_{s}}  \right) = \beta_0 + \beta_1 x_{1s}  + \beta_2 x_{2s}  + \beta_3 x_{3s}  + \beta_4 x_{4s},
\end{equation*}
where $\beta_i$ ($i=0,\dots,4$) are the parameters of interest. The design matrix is $\mathbf{D}=(\boldsymbol{X}_1, \dots, \boldsymbol{X}_4)$, where $\boldsymbol{X}_i$ is a column vector containing the $x_{is}$ values ($s=1,\dots,n$), with $x_{is}\in[-1,1]$. We define the following, independent prior distributions for each of the parameters of interest $\beta_i\sim U[a,b]$, $i=0,\dots,4$, where $a=(-3,4,5,-6,-2.5)$, and $b=(3,10,11,0,3.5)$.

We consider the cases where $n=6,10,24$, and $48$. The INSH algorithm for the logistic regression example is implemented in R (version 3.3.0).

\subsection{Code to Implement INSH}

\comment{
The online repository} \verb+http://www.github.com/DJPrice10/INSH_Code+ \comment{contains code to implement the INSH algorithm in MATLAB (Markovian death model), and R (PK model).}

\section{Results}
\label{section:results}

\subsection{Markovian Death Model}

We consider the optimal observation schedule when the number of observations permitted is $n=1,\dots4,6$ or $8$. The designs for $n=1,\dots,4$ observation times are compared to existing results, and $n=6$ and $8$ are reported as it was not computationally feasible using previous methods.

First, however, we provide a graphical demonstration of the INSH algorithm by considering two observation times for the death model. We choose to implement the INSH algorithm for $W=10$ iterations. We start with 100 randomly chosen designs across the feasible region, and retain the best $r_w=10$ and then $5$ designs (for 5 iterations each). At each generation, $m=3$ and then $6$ designs (for 5 iterations each), are sampled around each accepted design from the perturbation kernel, in order to sufficiently explore the space around each retained design. That is, we consider $m_w\times r_w=30$ designs at each iteration of the algorithm. The perturbation kernel in this example is a Normal distribution centred on the accepted design, with fixed standard deviation 0.1 for each design point (to allow reasonable exploration around each design point), zero covariances, and then truncated subject to the design constraints, \emph{i.e.}, $t_{i+1}-t_i>0$, $i=1,\dots,n-1$.

\begin{figure}[htbp]
	\begin{center}
		\subfloat[Generation 1.]{\includegraphics[width=0.425\linewidth]{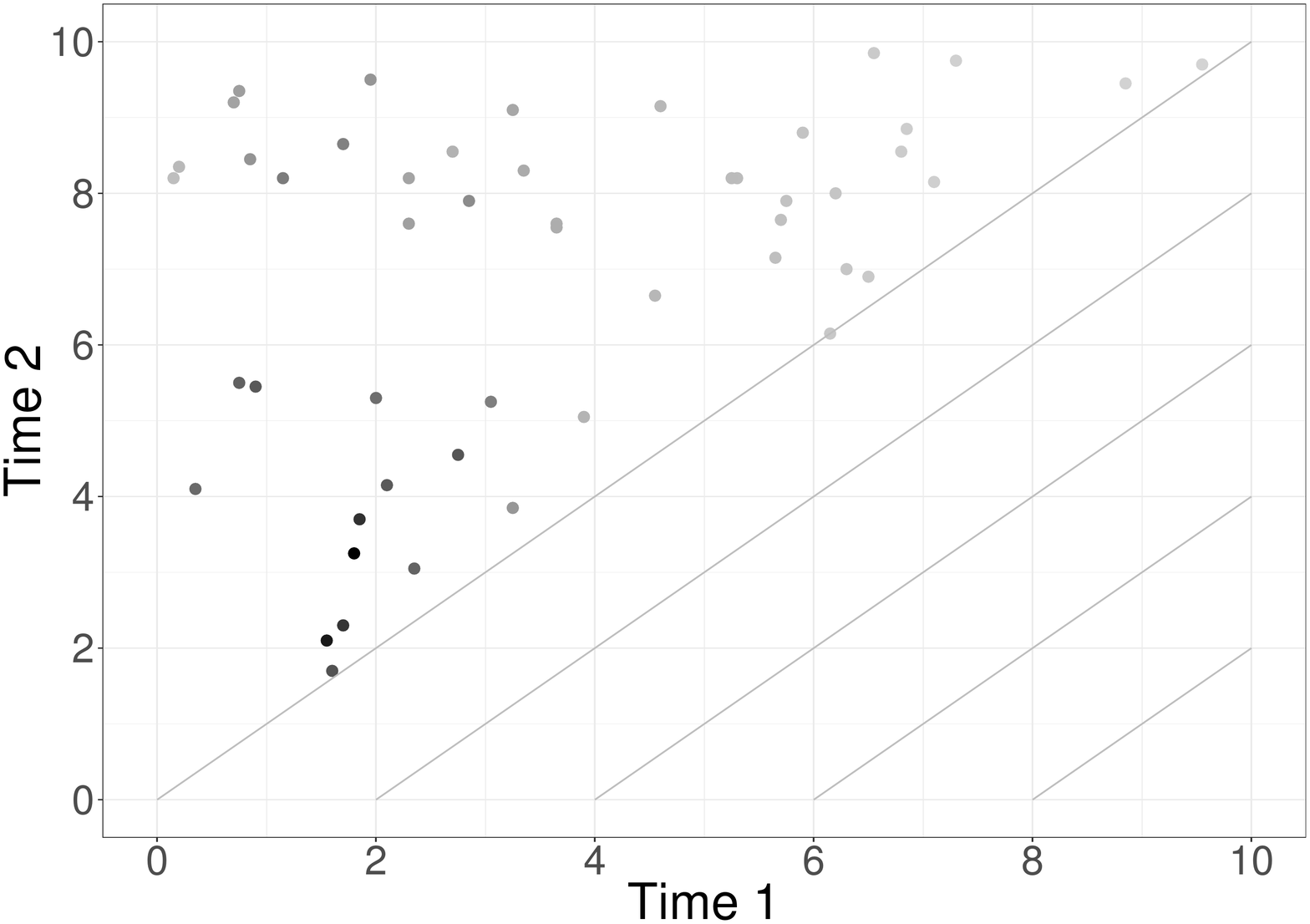}}
		\subfloat[Generation 2.]{\includegraphics[width=0.425\linewidth]{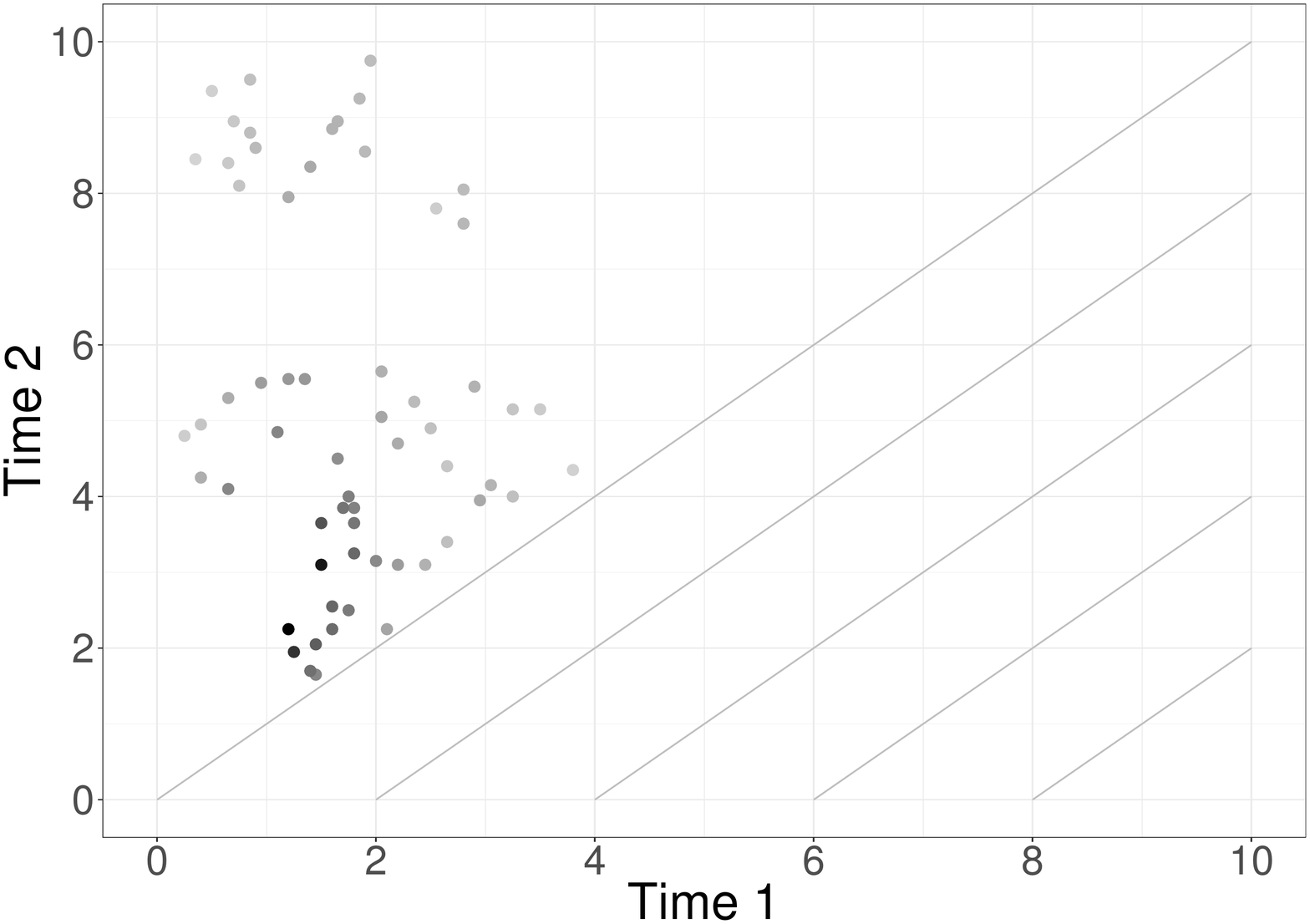}\label{INSH_G2}}\\
		\subfloat[Generation 3.]{\includegraphics[width=0.425\linewidth]{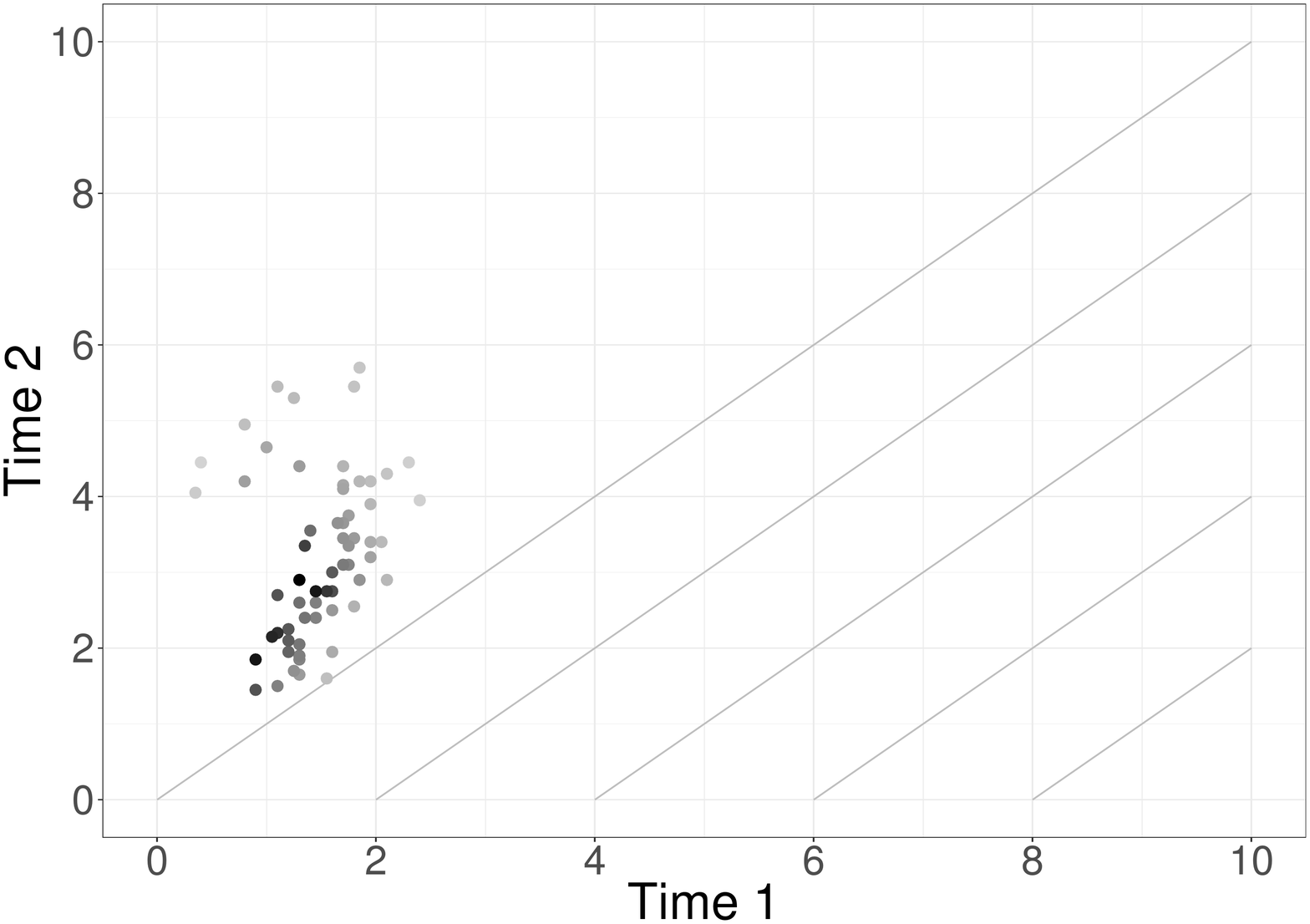}\label{INSH_G3}}
		\subfloat[Generation 4.]{\includegraphics[width=0.425\linewidth]{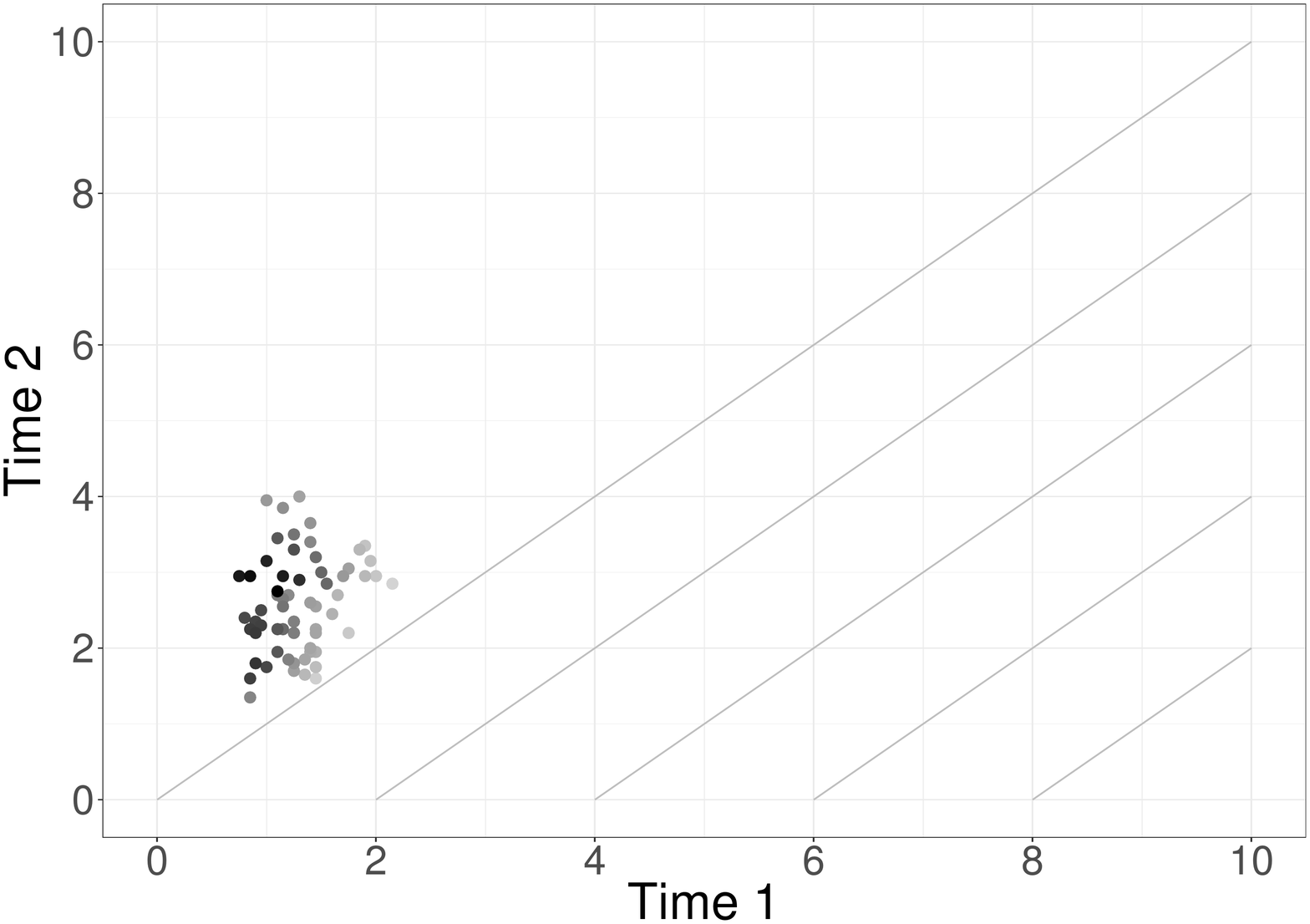}\label{INSH_G4}}\\
		\subfloat[Generation 5.]{\includegraphics[width=0.425\linewidth]{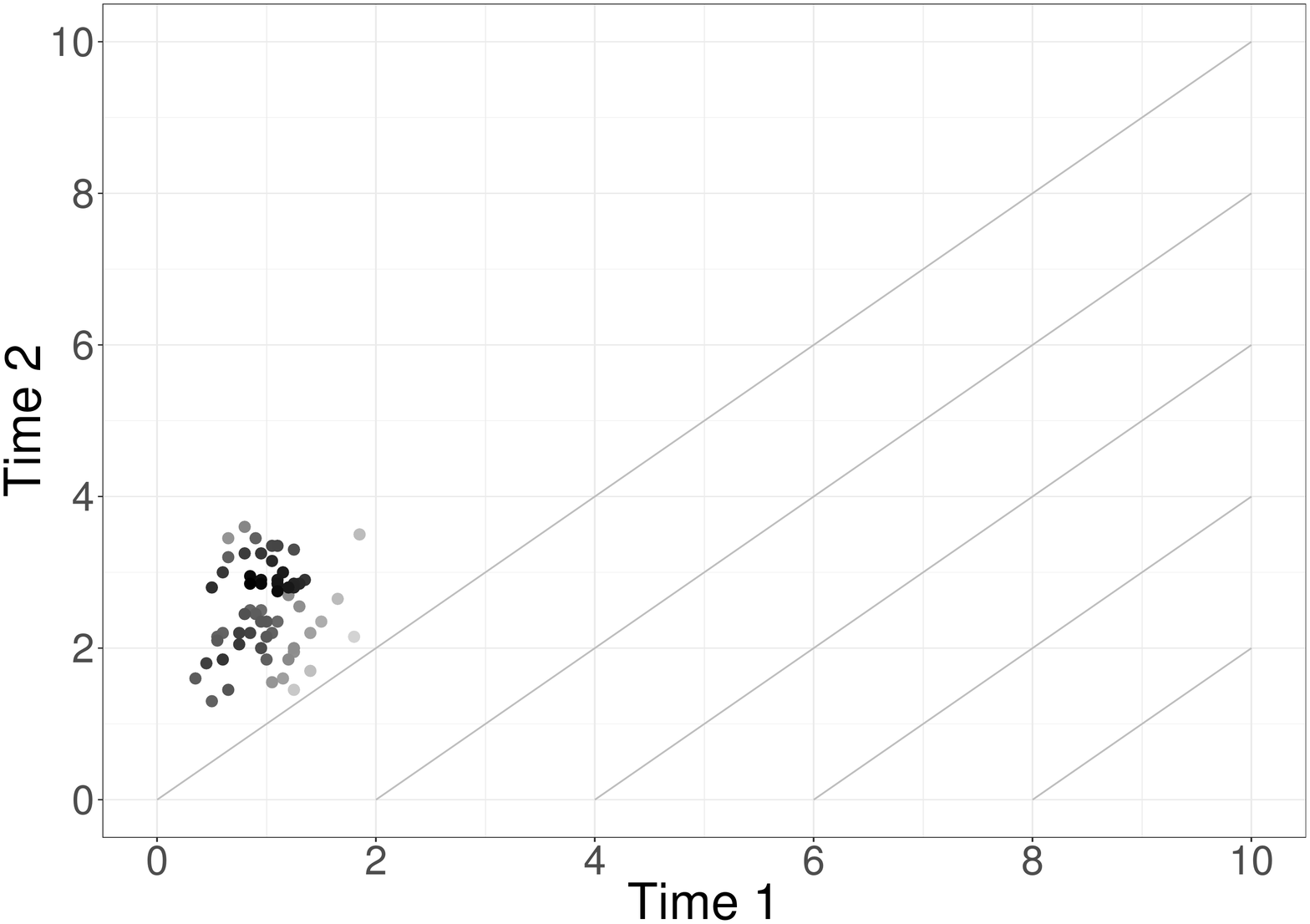}\label{INSH_G5}}
		\subfloat[Generation 6.]{\includegraphics[width=0.425\linewidth]{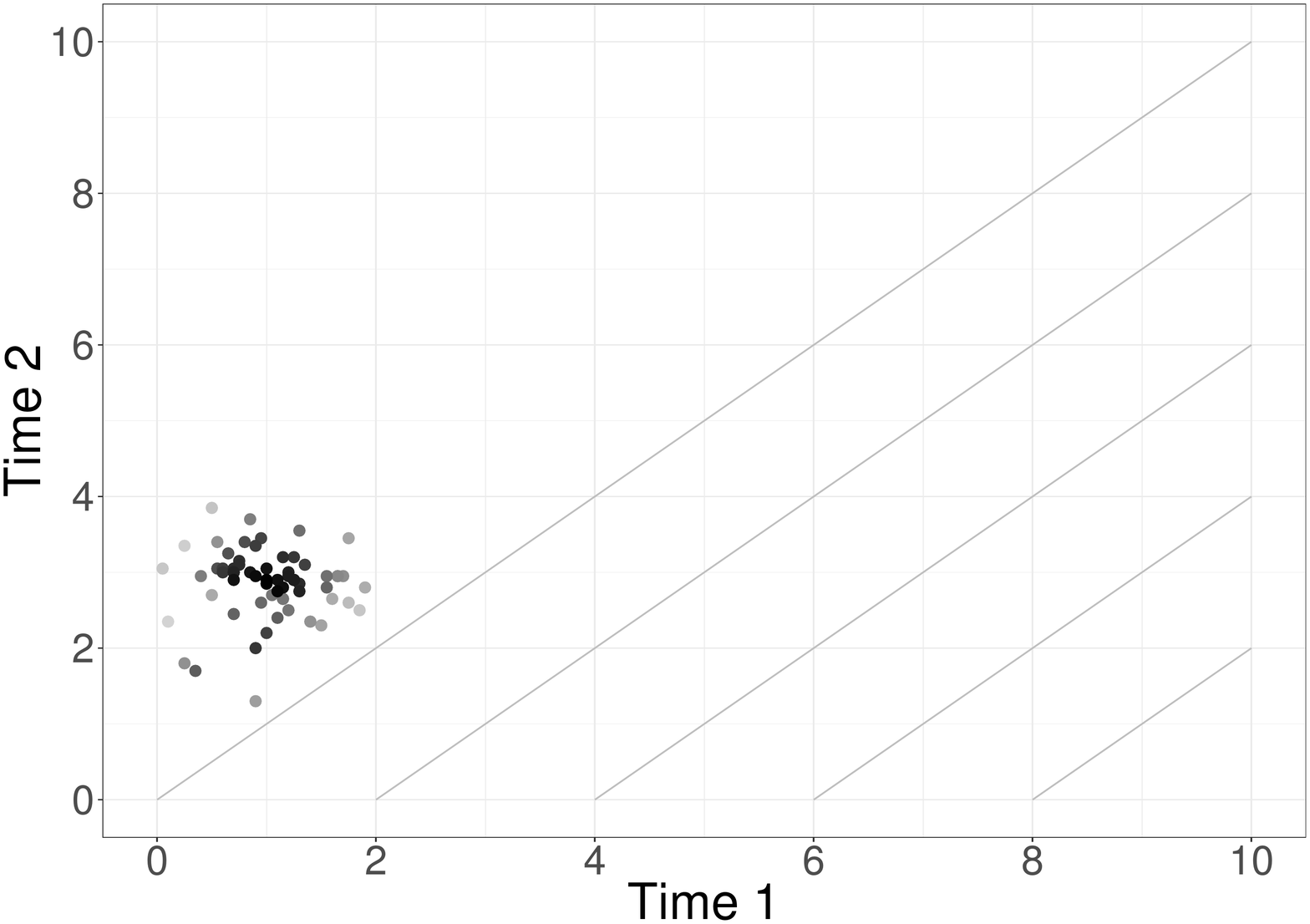}\label{INSH_G6}}\\
                \end{center}
                \caption{Demonstration of the design regions being considered by the INSH algorithm at each of the first six generations, and the convergence to regions of high utility. The shaded region corresponds to the infeasible design region (i.e., where $t_2<t_1$).} \label{deathmodel:stagesofINSH}
\end{figure}

\begin{figure}[htbp]
	\begin{center}
		\subfloat[]{\includegraphics[width=0.49\textwidth]{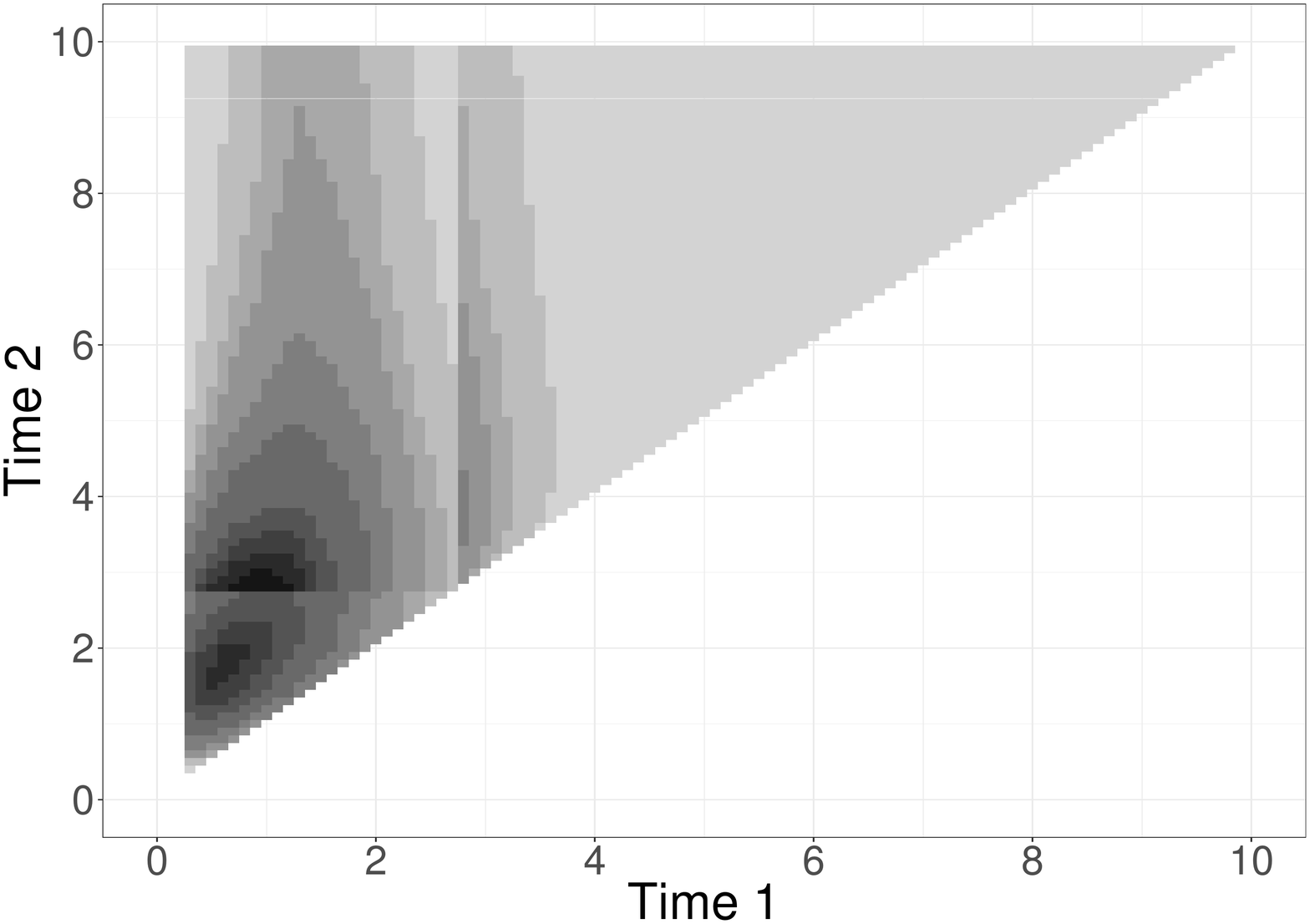}\label{deathmodel:fullutilitysurface}}
		\subfloat[]{\includegraphics[width=0.49\textwidth]{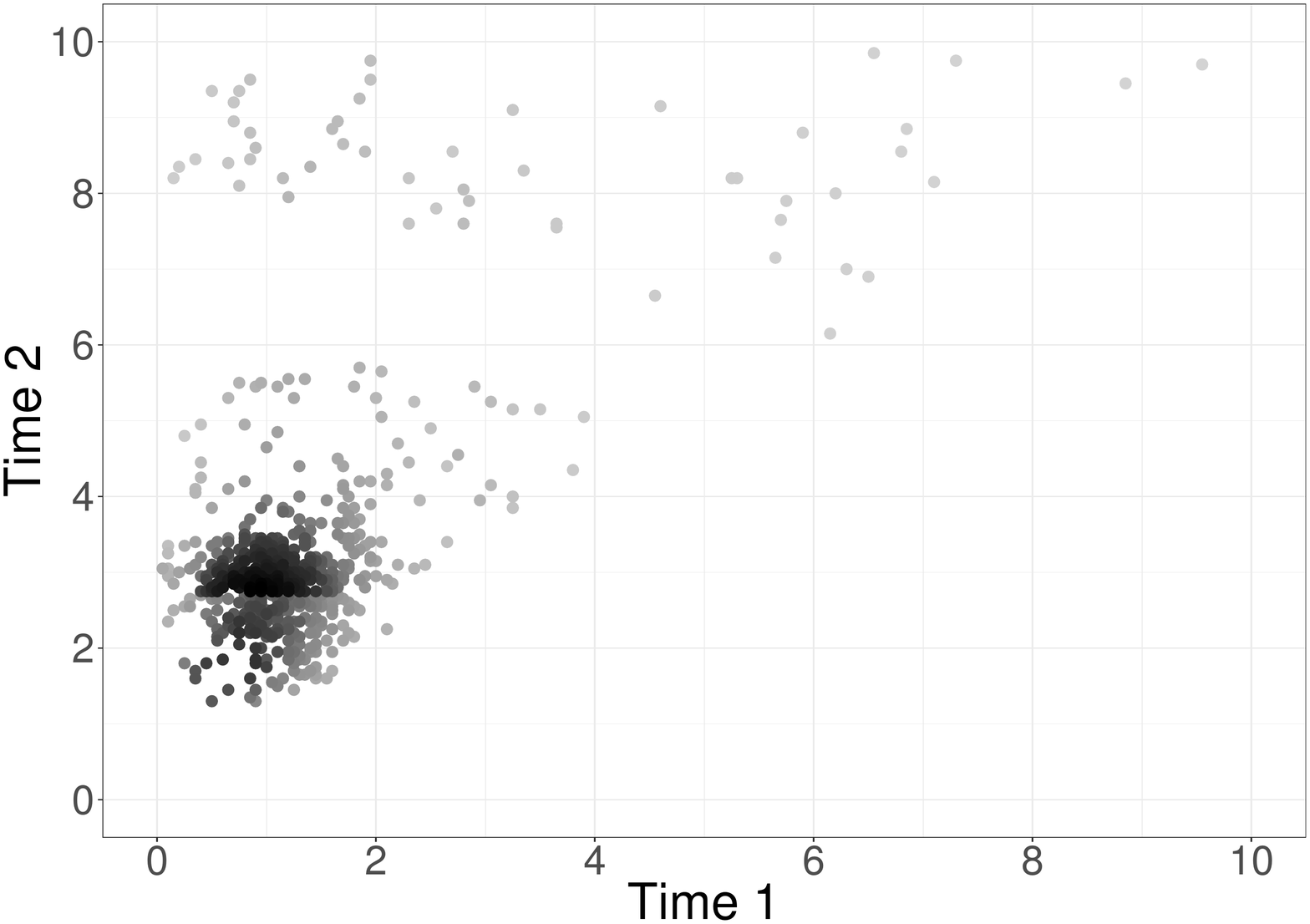}\label{PK_insh_final_samples}}
	\end{center}
	\caption{(a) Full utility surface for two observations of the death model evaluated on a grid using the ABCdE algorithm. (b) Samples from every generation of the INSH algorithm for two observations of the death model. In each figure, darker regions/points correspond to higher utility values.}
\end{figure}

Figure \ref{deathmodel:stagesofINSH} shows the progression of the INSH algorithm at each of the first six generations. For comparison, Figure \ref{deathmodel:fullutilitysurface} shows the full utility surface for the death model, evaluated using the ABCdE algorithm at all observation times across a grid with spacing 0.1, with $t_i\in[0.1,10]$. We can clearly see the optimal design is on a ridge at approximately (0.9, 2.8). There is also a region of high utility around (0.7, 2.0). Regions of low utility exist for very small $t_1$ (and in particular, $t_2>4$), or where both $t_1$ and $t_2$ are large (\emph{e.g.}, both above 3.5). In Figure \ref{deathmodel:stagesofINSH}, Generation 2 (Figure \ref{INSH_G2}) shows that regions of low utility are discarded early, and high utility regions are retained. Generations 2-6 (Figures \ref{INSH_G2}-\ref{INSH_G6}) demonstrate the convergence of the samples towards the region containing the optimal design. Generation 5 demonstrates the samples converging about the two ``peaks" observed in Figure \ref{deathmodel:fullutilitysurface} -- clearly demonstrating the ability to investigate multiple regions of high utility simultaneously. Figure \ref{PK_insh_final_samples} shows all design points considered throughout the INSH algorithm, with each point shaded by the utility value (darker corresponds to higher utility). The regions of high utility have been sampled more thoroughly. 


Online Resource B contains: box-plots illustrating the convergence of the sampled observation times towards the optimal, and the corresponding utilities towards the maximum in Figure S1; the optimal designs determined via INSH compared to the existing methods in Table S1, and; the corresponding INSH algorithm inputs in Table S2.

\subsection{Pharmacokinetic Model}
\label{section:PKresults}

Due to the physical constraints on the frequency at which sampling can be performed (at least 15 minutes apart), we restrict the designs such that $t_{i+1}-t_i \geq 0.25$, $i=1,\dots,14$. We sample designs from a multivariate-Normal perturbation kernel with fixed standard deviation 0.20, zero covariance, and truncated subject to the design constraints. The standard deviation was chosen such that one standard deviation was roughly the minimum distance between any two design points. The first generation of designs were sampled uniformly from the viable design space, $[0,24]$, subject to the constraints. As with the previous example, we specify a decreasing sequence of the number of retained designs $r_w$, and an increasing sequence of the number of sampled designs $m_w$.

In order to compare the run time of the ACE algorithm to the INSH algorithm, we implemented the ACE algorithm as detailed in \citet{Overstall:2017}, (\emph{i.e.}, running 20 instances of the ACE algorithm from the \verb+acebayes+ package in (embarrassingly) parallel fashion across four cores). On an iMac running OSX 10.11.4 with 4.0GHz Intel Core i7 processor and 32GB memory, this took 15.53 hours. We did not include the run time of the post-processing utility evaluation of the 20 candidate designs, 20 times each, in order to establish the overall optimal design, for reasons we state shortly. The ACE algorithm for this example in \citet{Overstall:2017} was performed 20 times from random initial conditions, each for a total of 20 iterations. Each iteration searches across each of the 15 dimensions of the design, and considers 20 candidate times to fit the Gaussian process. Thus, a total of $120{,}000$ designs are considered (\emph{i.e.}, utility evaluations) in the ACE algorithm, where $6{,}000$ of these utility evaluations are completed using significantly more Monte Carlo simulations. Specifically, the utility for the 20 candidate times used to train the Gaussian process are evaluated using $\tilde{B}=B=1{,}000$ Monte Carlo simulations, while the utility corresponding to the design with the proposed new observation time is evaluated using $\tilde{B}=B=20{,}000$.

The advantage of the INSH algorithm is in its ability to consider a large number of designs in multiple regions simultaneously, and so it is sufficient to use less effort to evaluate the utility of each design, as a noisy estimate of the utility will have less influence on the output of the algorithm. Hence, we used $\tilde{B}=B=5{,}000$ for the evaluation of the utility of each design, which was completed in parallel on four cores (using \verb+foreach+ and \verb+doParallel+ packages in R), on the same machine as stated above. In particular, at each generation of the algorithm, the calculation of the utilities of the designs in the current wave was split across the number of available cores (\emph{i.e.}, 1/4 of the required utility calculations were allocated to each core). We ran the INSH algorithm for $W=60$ iterations, with $1{,}200$ randomly generated initial designs. At each iteration, we retained the ``best" 150, 75, 50,  25, and 10 designs, and proposed two, four, six, 12 and 30 new designs around each accepted design, for 12 iterations of each combination -- maintaining consideration of 300 designs at each iteration, while regularly increasing the exploitation and reducing exploration. These values of $r_w$ and $m_w$ were chosen such that earlier generations of the algorithm retained a reasonable number of designs -- thus, not excluding regions of the design space too quickly -- and as the algorithm progressed, focussed computational effort on high-utility regions of the design space. Given the larger dimension of the design in this example (compared to the death model), we chose to sample a large number of designs around each retained design in later generations of the algorithm, in order to sufficiently explore the design space in proximity to the optimal.  

This run of the INSH algorithm took 2.23 hours (approximately 7 times faster than the ACE algorithm). Having obtained the designs and utility evaluations from the INSH algorithms, we perform the same post-processing utility evaluation on the 20 best considered designs, with 20 evaluations of the utility of each design with $\tilde{B}=B=20{,}000$, in order to identify the overall optimal. The total number of designs considered by the INSH algorithm with this selection criteria is approximately: $(\text{No.\ initial designs}) + (W-1)\times r_w\times m_w=1200 + (60-1)\times 300 = 18{,}900$ -- that is, the number of initial designs, plus how many were retained at each generation multiplied by the number that were sampled around each retained design. In practice, this number is often slightly higher, as the $r_w^{th}$ ranked design can be a tie, and the optimal design is re-introduced into the set of designs being considered if it occurred in a previous generation (this run of the INSH algorithm actually considered $19{,}428$ designs).

\begin{figure}[htbp]
	\begin{center}
		\includegraphics[width=0.65\linewidth]{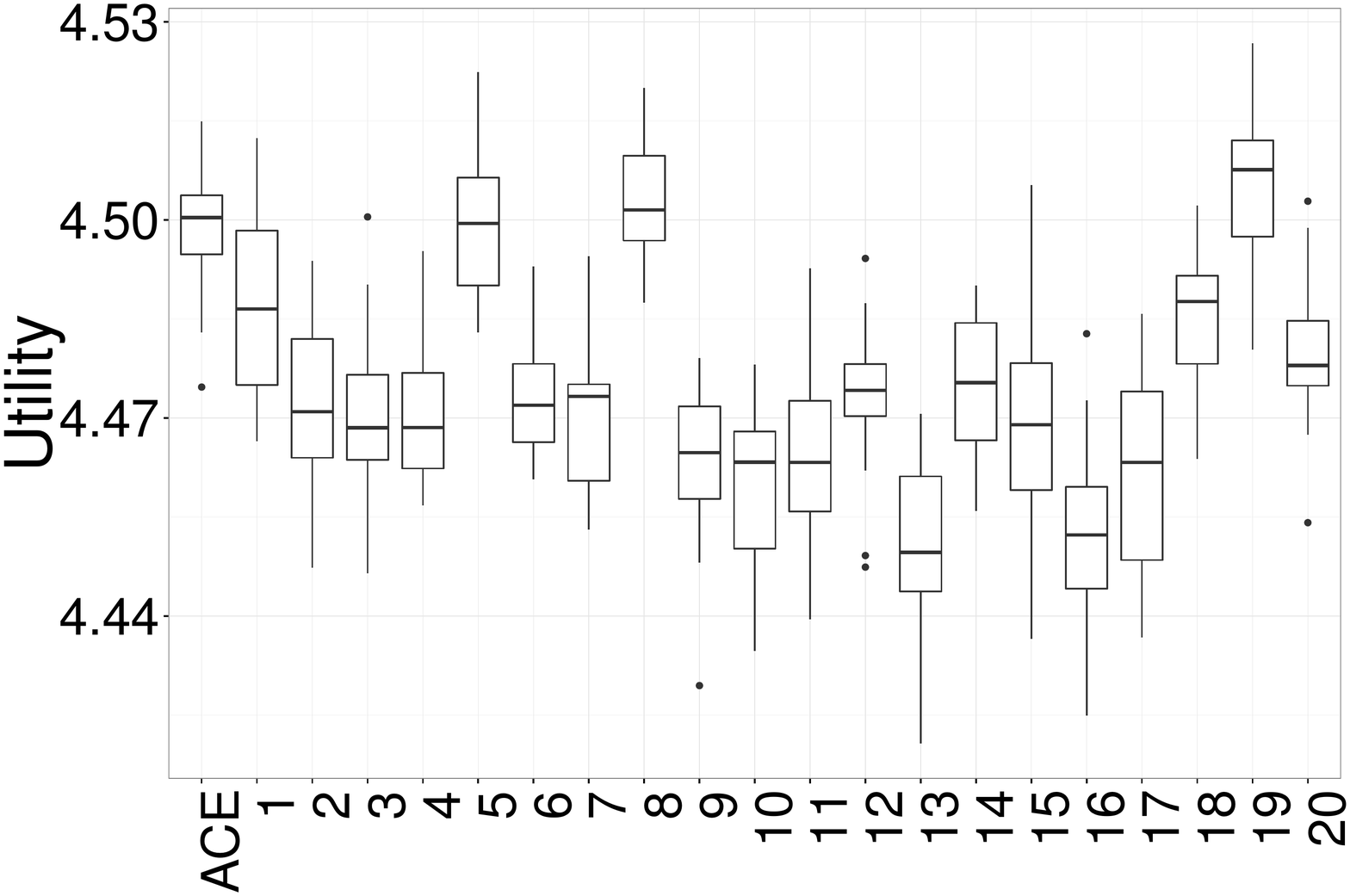}
	\end{center}
	\caption{Box-plots of the utility for the optimal design found by the ACE algorithm, compared to the top 20 designs considered by the INSH algorithm. The utility of each design is evaluated 20 times, using $\tilde{B}=B=20{,}000$ Monte Carlo simulations.} \label{PK_all_INSH_designs}
\end{figure}

Figure \ref{PK_all_INSH_designs} shows box-plots of the 20 utility evaluations for each of the 20 best designs that were considered by the INSH algorithm, compared to the same number of evaluations of the ACE optimal design reported in \verb+optdescomp15sig()+ in the \verb+acebayes+ package (each utility evaluation using $\tilde{B}=B=20{,}000$). We can see from this figure that there are three designs (5, 8, 19), that perform similarly well to the design found using the ACE algorithm. Online Resource C contains: these three designs from INSH in Table S3; summaries of the utilities for the top 20 designs evaluated by INSH in Table S4; box-plots demonstrating the convergence of the INSH algorithm to the optimal region in Figure S3, and; a comparison of the ACE and INSH optimal designs performance with regards to inference in Figure S4 (in particular, the posterior variance and bias in posterior mode).

\subsubsection{Sampling Windows}

The population-based approach of INSH means that we retain a large number of designs with high utility. We use these ``best'' designs to construct the sampling windows for each sampling time, similar to the approach of \citet{McGree:2012}. \citet{McGree:2012} used percentiles of the designs evaluated once a stopping-criteria has been reached in their algorithm to form the sampling windows -- we choose a fixed number of ``best'' designs to form the windows. Given the windows, those implementing the design can choose observation times from these windows, ensuring that the physical constraint, $t_{i+1}-t_i\geq0.25$, is satisfied.

\begin{figure}[htbp]
	\begin{center}
		\subfloat[]{\includegraphics[width=0.49\textwidth]{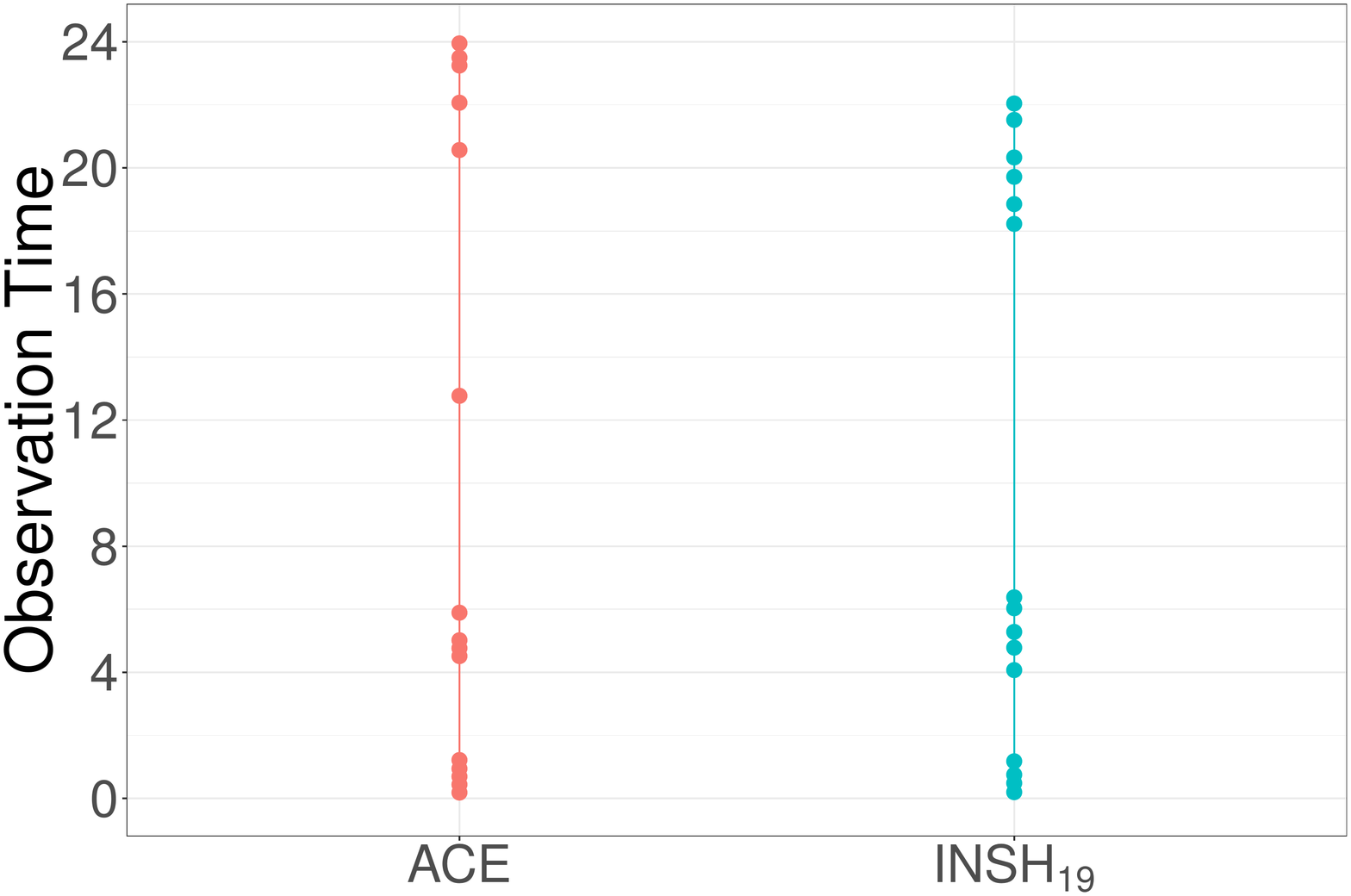}\label{PK_insh_v_ace} \label{PK_all_plots:opt_comparison}}
		\subfloat[]{\includegraphics[width=0.49\textwidth]{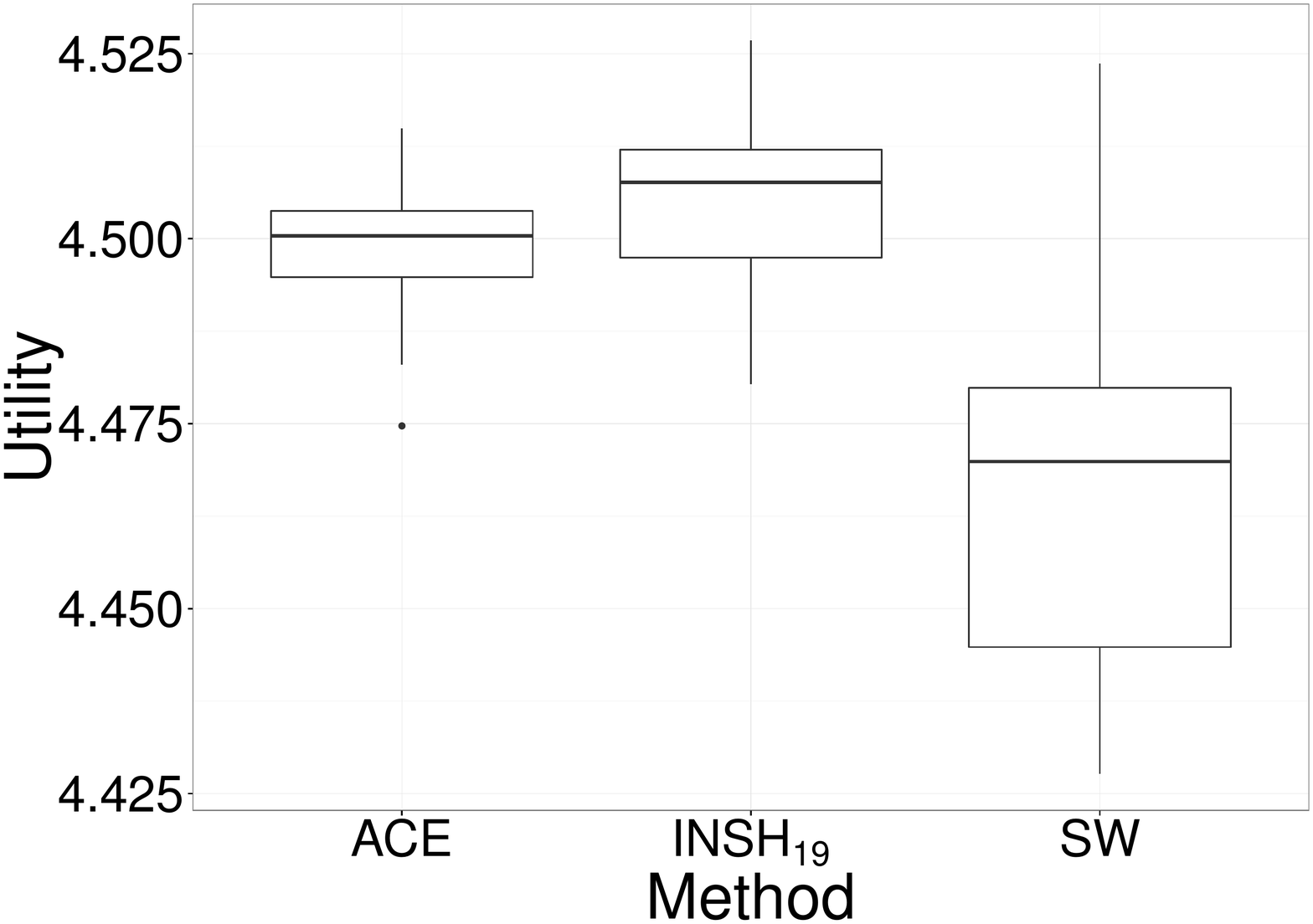}\label{PK_utils} \label{PK_all_plots:util_comparison}} \\
		\subfloat[]{\includegraphics[width=0.49\textwidth]{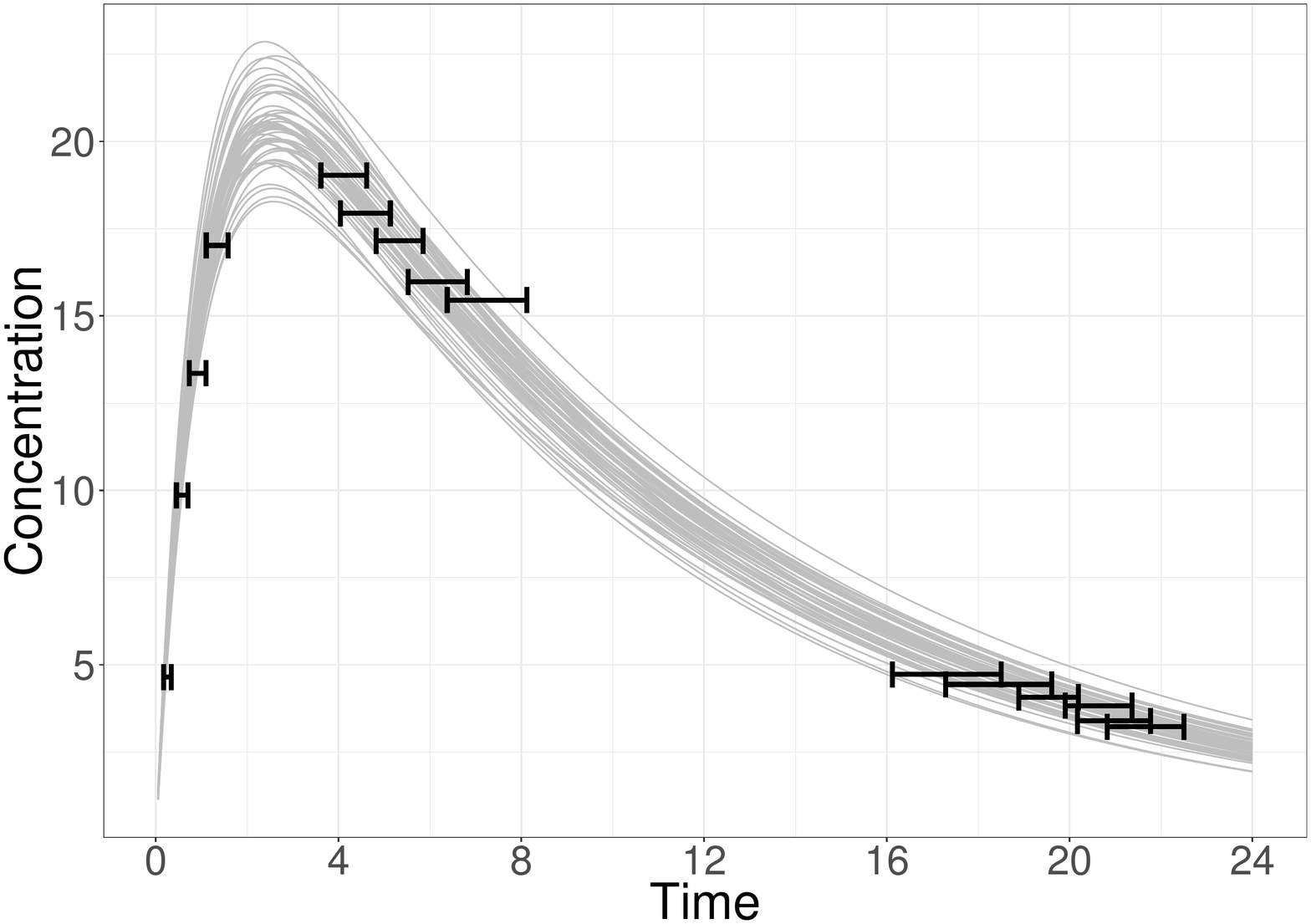}\label{PK_sw_lineplots} \label{PK_all_plots:sw_range}}
		\subfloat[]{\includegraphics[width=0.49\textwidth]{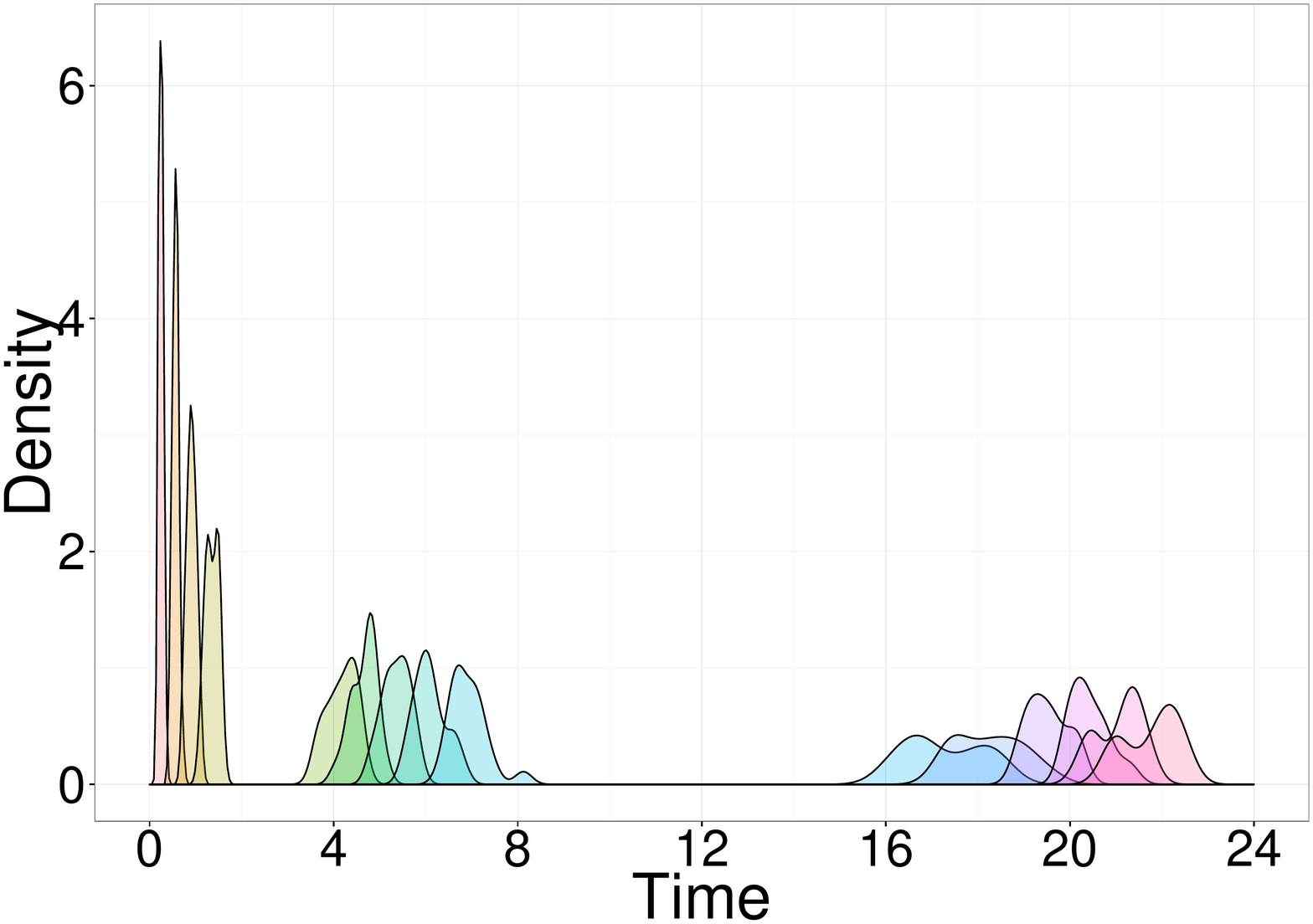}\label{PK_sw_util} \label{PK_all_plots:sw_density}}
	\end{center}
	\caption{(a) Comparison of the optimal designs from the ACE and INSH methods. (b) Boxplots of 20 utility evaluations for the ACE and INSH optimal designs, and the INSH sampling windows designs, using $\tilde{B}=B=20000$. (c) Sampling windows for each observation time obtained from the 20 designs corresponding to the highest utilities found during the INSH algorithm, plotted over 50 realisations of the PK model simulated at parameter values randomly drawn from the prior distribution. The error bars show the sampling window range. (d) Density plot of the sampling windows for each observation time. 
	} \label{PK_all_plots}
\end{figure}

As an example of this process, we arbitrarily consider the top 20 designs from the output of the INSH algorithm for the PK example, and form sampling windows as the range of values considered at each observation time for these ``best'' designs. Alternatively, one could consider all designs that were within some percentage of the utility corresponding to the maximum, or, use a weighting based on the average utility for each design to approximate a distribution for each sampling time which could subsequently be sampled. In order to construct the sampling window designs, we ``bootstrap" an observation schedule by randomly selecting each of the 15 sampling times (with equal probability), from the 20 candidate observation times, subject to the constraints. A new design is sampled for each of the 20 utility evaluations to demonstrate the range of potential outputs from this approach. Figures \ref{PK_all_plots:sw_range} and \ref{PK_all_plots:sw_density} show the INSH sampling windows for each observation time. Figure \ref{PK_all_plots:opt_comparison} shows the optimal observation schedules evaluated using the ACE and INSH algorithms. Note that the optimal design returned from the INSH method corresponded to the $19^{th}$ highest utility value from the original output of the INSH algorithm (\emph{i.e.}, using $\tilde{B}=B=5{,}000$). It was deemed the optimal design as it corresponded to the highest mean utility, from 20 utility evaluations using $\tilde{B}=B=20{,}000$ (Figure \ref{PK_all_INSH_designs}). Figure \ref{PK_all_plots:util_comparison} shows box-plots of 20 utility evaluations (using $\tilde{B}=B=20{,}000$) for the ACE and INSH optimal designs, and the 20 randomly selected designs from the sampling windows. Note that the average efficiency of the sampling windows designs compared to the INSH optimal design is 99.07\%.

\subsection{Logistic Regression in Four Factors}

We implement the INSH design for $n=6,10, 24$, and $48$, and compare the utility for the best performing design found via INSH to those reported from ACE. INSH is implemented in the same way as for the PK example. We step down the value of $r_w$, and increase the value of $m_w$ as $w$ increases, such that early iterations are geared towards exploration, while later iterations are focussed on exploitation. Table S5 in Online Resource D contains the values of $m_w$ and $r_w$ that are used for the INSH algorithm. In this example, we utilise a uniform perturbation kernel with a fixed-width. In order to further increase exploitation as the algorithm progresses, we step down the width of this proposal distribution in line with the change in $r_w$ and $m_w$. For each example, we retained a reasonable number of designs in early generations of the algorithm in order to exclude less of the design space, and increase computational effort on regions of high utility in later generations, as with the PK example. Given the high-dimensional design space for the $n=48$ example, we sampled more designs around each retained design at each generation (\emph{i.e.}, larger $m_w$), in order to better explore the design space. Furthermore, as we are considering very high-dimensional design spaces, we run the risk of randomly stepping in the wrong direction from a given design -- and one cannot feasibly explore satisfactorily around each design point. To avoid the potential for stepping in a poor direction and not being able to get back to a region of potentially high utility, we slightly alter our acceptance step to consider the best $r_w$ designs out of the current iteration, \emph{and} those that were accepted in the previous iteration. This way, should we move from a region of high utility to a region of low utility through a poorly proposed design, we are able to essentially take one-step back, and propose a new design from the previous design. 

\begin{figure}[htbp]
	\begin{center}
		\includegraphics[width=0.85\textwidth]{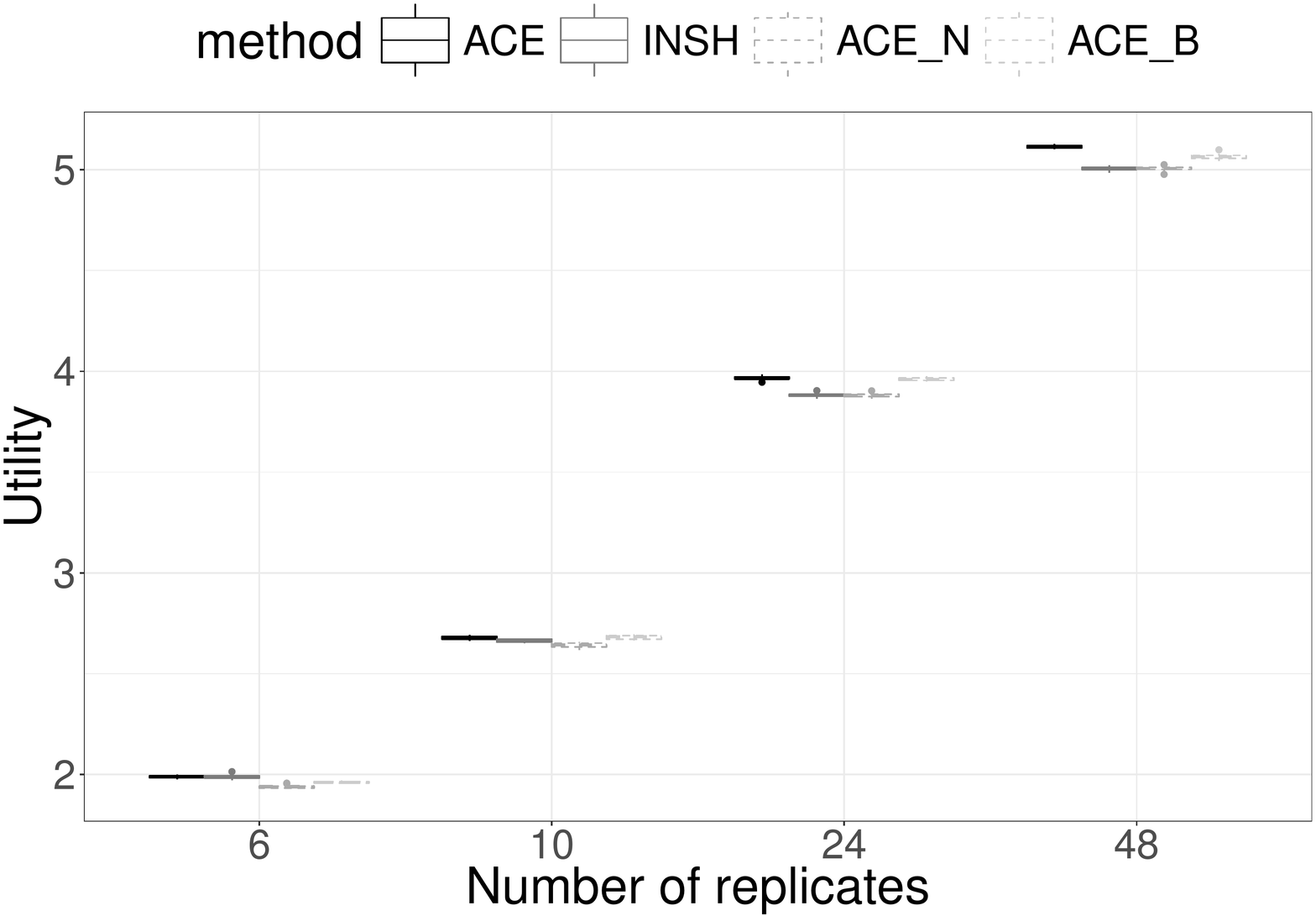}
	\end{center}
	\caption{(a) Box plots of 20 utility evaluations for the optimal designs found by each of the ACE, INSH (solid), ACE$_N$, and ACE$_B$ (dashed) methods, for $n=6,10,24$ and $48$.}
	\label{LR_util}
\end{figure}

\comment{
Initially, our approach was to run INSH for the same computation time as ACE in order to establish the utility of the best designs found via each method. However, the INSH algorithm was unsuccessful in converging to designs with the same utility of those found via ACE for scenarios with a larger design space (INSH designs contained approximately 98.5\% of the utility relative to the ACE designs). Instead, we ran INSH for a reasonable set of parameter values to determine near-optimal designs in a reasonable amount of computation time, and then adjusted the input parameters for ACE to run for the same computation time as INSH. In particular, we reduced the number of iterations of the two phases of the algorithm ($N_I,N_{II}$), or the amount of effort used in evaluating the utility in both training the Gaussian process, and choosing to update a coordinate ($B$). We denote these implementations of ACE as ACE$_N$ and ACE$_B$, respectively. Figure \ref{LR_util} shows box-plots of the utility evaluated at designs found via each implementation of INSH and ACE. For each $n$, the designs found via ACE$_N$ appear to perform as well as those found via INSH, however, ACE$_B$ designs perform better than those found via INSH for larger design spaces (i.e., $n>10$). We note that for $n=6$, and $10$, INSH finds similarly-performing designs to ACE for the same computation time. In Online Resource D: Tables S6-S9 show the designs found via the INSH algorithm; Figures S5-S8 show box-plots of the utility of each design considered at each iteration of the INSH algorithm for each $n$; Table S10 shows the mean and 2.5-97.5\%-percentiles of the designs found by each of the ACE, INSH, ACE$_N$ and ACE$_B$ algorithms, and; Table S11 shows the input parameters for ACE$_N$ and ACE$_B$ to achieve the same computation time as INSH for each $n$.}

\comment{
We do not believe that the discrepancy in performance of the designs found via the ACE and INSH methods is due solely to the increased dimension of the design space. Rather, we believe that it is a combination of the high-dimensional setting rendering the perturbation step less effective, and that the optimal design in each case resides on the boundary. In considerably high-dimension problems, the perturbation step results in designs considered by INSH routinely moving away from some, or all, of the boundary values that would otherwise result in a more informative design. In other words, the resampling approach of INSH means that it is highly unlikely to stay at a large number of boundary values simultaneously. Conceptually, one can see that designs that reside away from the boundary values can be approached from any direction, whereas boundary values can only be approached from, in a loose sense, ``one direction". In these examples, the designs contain many boundary values -- for n=6, 10, 24 and 48 the optimal designs via ACE contain 16/24, 28/40, 70/96, and 143/192 boundary values (i.e., -1 and 1's), respectively. We acknowledge that the optimal design existing on a boundary is a common feature of multi-factor experiments such as this, and that this example has highlighted a shortcoming of the INSH algorithm. However, it is the authors belief that INSH is still a suitable method for high-dimensional problems, however we are unable to demonstrate this with an existing high-dimensional design problem at this stage. }

\comment{
Given that INSH out-performs ACE for small-moderate design spaces -- as demonstrated for both the PK example and LR example (for a fixed computation time) -- we propose that INSH is a suitable, computationally-efficient alternative to the ACE algorithm for up to 40-dimensional design spaces (i.e., corresponding to $n=10$ in this example). Otherwise, for truly high-dimensional design problems (i.e., more than 40-dimensions), the authors propose that ACE is implemented given it has been shown to perform well in these scenarios.
}

\section{Discussion}
\label{section:discussion}

In this paper, we have considered three common types of statistical model: a Markovian death model, a one-compartment PK model, and a four-factor logistic regression model. Our results for the death model provide a simple demonstration of the efficacy of the proposed INSH algorithm, and gives equivalent answers to previously applied methods. The PK model allowed us to consider a larger design space, and show that the INSH algorithm is able to return designs that marginally outperformed those found using the ``gold-standard" ACE algorithm, in considerably less computation time -- illustrating that for moderate-size design spaces, INSH is more computationally efficient than ACE. We also showed the simple extension to the INSH algorithm that allows the construction of sampling windows. Finally, the logistic regression example provided an example of a truly high-dimensional design space. We were unable to find designs that performed as well as those found via ACE, however, we were able to demonstrate that INSH provides a suitable, computationally-efficient approach for up to approximately 40-dimensions -- which encompasses a large range of experimental design problems. Alternative examples of considerably high-dimension with optimal designs residing away from the boundary should be considered in order to demonstrate the performance of INSH relative to ACE in these instances.

We have not provided a proof that the INSH algorithm will converge to the optimal design, however, one can see that in the limit (\emph{i.e.}, $W\rightarrow\infty$, $\alpha_0=0$ and $\alpha_w\rightarrow1$ as $w\rightarrow\infty$, and sufficiently large $m$), the INSH algorithm will identify the optimal design. 
However, as with many optimisation routines, the aim of this algorithm is to find near-optimal designs in a computationally feasible amount of time. Thus, practical algorithm inputs must be chosen, which may not guarantee convergence to the optimal solution. However, this trade-off is apparent in a number of existing optimisation routines -- for example, simulated-annealing, cross-entropy, and genetic algorithms all have the potential to converge to local, rather than global, optima.

The INSH algorithm we have presented here is quite general, and there exist many aspects of the algorithm which can be explored in order to improve the efficiency of the algorithm for different optimisation problems. For example: one could update the perturbation kernel based on the correlation/covariance that exists between design parameters of the same design, or; randomly incorporate a sample in a region of the design space that has either not been considered previously, or was dismissed earlier in the algorithm, in order to increase exploration of the design space and maximise the chance of obtaining the optimal design. Another important consideration will be to provide some general rules regarding the choice of algorithm inputs for a particular utility surface, or magnitude and dimension of design space. For example, the initial samples could be used to approximate some characteristics of the utility surface, and provide some insight into sensible choices of the inputs for the algorithm. While we did not consider it here, increasing the number of utility evaluations which form the approximate expected utility could also be increased as the algorithm progresses, \emph{i.e.}, specify a sequence for $\tilde{B}$ and $B$ in the SIG utility evaluation -- ensuring more effort is spent evaluating a more precise estimate of the utility in regions near to the optimal design. 

While we have added commentary around our choice of parameters for each example, we note that many of the parameters are problem-specific, and require the user to specify sensible values based on their understanding of the system and design space. As with other stochastic optimisation routines, some trial-and-error may be required in order to choose suitable INSH input values for different problem types. Given the drastic increase in computational-efficiency of the INSH algorithm for small-to-moderately sized design problems, the authors believe that one could be very cautious with some parameter choices (\emph{e.g.}, choose a large number of accepted designs, and number of newly sampled designs), in order to ensure satisfactory exploration of the design space, and obtain designs more efficiently than existing algorithms.

\section*{Acknowledgements}
DJP acknowledges the support of the BBSRC (BB/M020193/1) awarded to Olivier Restif. JVR acknowledges the support of the ARC (Future Fellowship FT130100254; CoE ACEMS) and the NHMRC (CRE PRISM$^2$). The authors would like to sincerely thank the reviewers whose comments greatly helped shape and improve this manuscript.



\newpage
\section*{References}


%
%
%

\newpage
\setcounter{figure}{0} 
\renewcommand\thefigure{S\arabic{figure}}
\renewcommand\thetable{S\arabic{table}}

\section*{Appendix A} \textit{Existing Algorithms}

Algorithm \ref{ourabc} describes the ABC algorithm used by ABCdE and the INSH algorithm (for the death model) to evaluate the posterior distribution.
\begin{algorithm}[htbp]
\caption{ABC Algorithm: Fixed tolerance}\label{ourabc}
\begin{algorithmic}[1]
\Require Observed data $\boldsymbol{x}$, simulated data $\boldsymbol{y}=(\boldsymbol{y}^1,\dots,\boldsymbol{y}^N)$, corresponding parameter values $\boldsymbol{\theta}^i, i=1,\dots,N$, and tolerance $\epsilon$.
	\State Evaluate discrepancies $\rho^i = \rho(\boldsymbol{x}, \boldsymbol{y}^i)$, creating particles $\{ \boldsymbol{\theta}^i, \rho^i \}$ for $i=1,\dots,N$.
	\State Using the posterior sample of parameters $\boldsymbol{\theta}^i$ such that $\rho^i<\epsilon$, evaluate utility.
\Ensure Utility for current design, having observed $\boldsymbol{x}$, $U(d,\boldsymbol{x})$.
\end{algorithmic}
\end{algorithm}

Algorithm \ref{ABCdEalgorithm} describes the ABCdE algorithm of \citet{Price:2016}, to evaluate the optimal Bayesian experimental design.
\begin{algorithm}[H]
\caption{ABCdE Algorithm}\label{ABCdEalgorithm}
\begin{algorithmic}[1]
	\State Choose grid over the parameter space for the discrete estimate of the utility, number of simulations $N_{pre}$, and tolerance $\epsilon$.
	\State Sample $N_{pre}$ parameters $\boldsymbol{\theta}$ from $p(\boldsymbol{\theta})$. 
	\State For each of the $N_{pre}$ parameters, and under every design $d$ in the design space $\mathcal{D}$, simulate process and store $X_{N_{pre}\times |\mathcal{D}|}(\boldsymbol{\theta}, d)$. \label{abcde_algorithm_line3}
	\For{$i=1$ to $|\mathcal{D}|$}
	\State \parbox[t]{0.925\linewidth}{Consider the unique rows of data $Y(\boldsymbol{\theta}, d^i) = \text{ unique}(X(\boldsymbol{\theta}, d^i))$.\\ \emph{Note: We let $K^i$ be the number of such unique data, and $n_{k^i}$ be the number of repetitions of the ${k^i}^{th}$ unique data, for $k^i=1,\dots,K^i$}.\label{abcdealg:uniquedata}\vspace{4pt}}
		\For{$k^i=1$ to $K^i$}
		\State \parbox[t]{0.925\linewidth}{Pass `observed data' $\boldsymbol{y}^{k^i}=[Y(\boldsymbol{\theta},d^i)]_{k^i}$, `simulated data' $X(\boldsymbol{\theta},d^i)$, $N_{pre}$ sampled parameters, and tolerance $\epsilon$ to Algorithm \ref{ourabc}, and return contribution $U(\boldsymbol{y}^{k^i},d^i)$ to the expected utility, for ${k^i}^{th}$ unique datum (`observed data') and $i^{th}$ design\label{abcdealg:createposterior}.}
		\EndFor
	\State {Store $u(d^i) = \frac{1}{N_{pre}} \sum_{k^i} {n_{k^i}}  U(\boldsymbol{y}^{k^i}, d^i)$; the average utility over all parameters and data for design $d^i$. \label{abcde_algorithm_line9}}
	\EndFor
	
	\Ensure The optimal design $d^* = \underset{d\in\mathcal{D}}{\text{argmax}}(u(d))$.
\end{algorithmic}
\end{algorithm}

Algorithm \ref{mullersalgorithm} details the MCMC algorithm for determining Bayesian optimal designs proposed by Muller [1999].
\begin{algorithm}[htbp]
\caption{MCMC with stationary distribution $h(\boldsymbol{\theta},\boldsymbol{x},d)$, Muller [1999]}\label{mullersalgorithm}
\begin{algorithmic}[1]
\Require Number of samples $m$, prior distribution of model parameters $p(\boldsymbol{\theta})$, and proposal density $q(\cdot)$.
	\State Choose, or simulate an initial design, $d^1$.
	\State Sample $\boldsymbol{\theta}^1\sim p(\boldsymbol{\theta})$, simulate $\boldsymbol{x}^1\sim p(\boldsymbol{x}\mid \boldsymbol{\theta}^1, d^1)$, and evaluate $u^1=U(\boldsymbol{\theta}^1, \boldsymbol{x}^1, d^1)$.\label{mulleralgorithm:initial}
	\For{$i=1:m$} 
	\State Generate a candidate design, $\tilde{d}$, from a proposal density $q(\tilde{d} \mid d^i)$.
	\State Sample $\tilde{\boldsymbol{\theta}}\sim p(\boldsymbol{\theta})$, simulate $\tilde{\boldsymbol{x}}\sim p(\boldsymbol{x}\mid \tilde{\boldsymbol{\theta}}, \tilde{d})$, and evaluate $\tilde{u}=U(\tilde{\boldsymbol{\theta}}, \tilde{\boldsymbol{x}}, \tilde{d})$. \label{mulleralgorithm:simdata}
	\State Calculate,
		\begin{align}
\alpha	&=\min\left\{ 1, \frac{\tilde{u}\ q(d^i\mid \tilde{d})}{u^i\  q(\tilde{d} \mid d^i)} \right\}. \notag
		\end{align}
		\State Generate $a\sim U(0,1)$
	\If{$a<\alpha$} 
			\State Set $(d^{i+1}, u^{i+1}) = (\tilde{d}, \tilde{u})$
	\Else{}
			\State Set $(d^{i+1}, u^{i+1}) =(d^i, u^i)$
		\EndIf
	\EndFor\\
	\Ensure Sample of $m$ designs, $d$.
\end{algorithmic}
\end{algorithm}

\newpage
\section*{Appendix B} 

\textit{Markovian Death Model ABC Choices}

{We provide the parameter choices for the ABC algorithm used to evaluate the approximate posterior distributions when evaluating the utility for the Markovian death model example. 
Prior to running the ABC algorithm (Algorithm \ref{ourabc}), we sample $N=50,000$ parameter values from the prior distribution, and simulate data corresponding to each under each design. For each of 1, 2, 3, 4, 6, and 8 observation times, we use a tolerance of 0.25, 0.50, 0.75, 1.00, 1.50, 1.50, respectively.}

{We note however, that these choices are problem specific, and suggest that researchers undertake a pilot-study in order to determine sensible parameter choices, as one would do prior to using ABC for inference.}

\textit{Markovian Death Model Results}

\begin{figure}[H]
	\begin{center}
		\subfloat[First observation time.]{\includegraphics[width=0.45\linewidth]{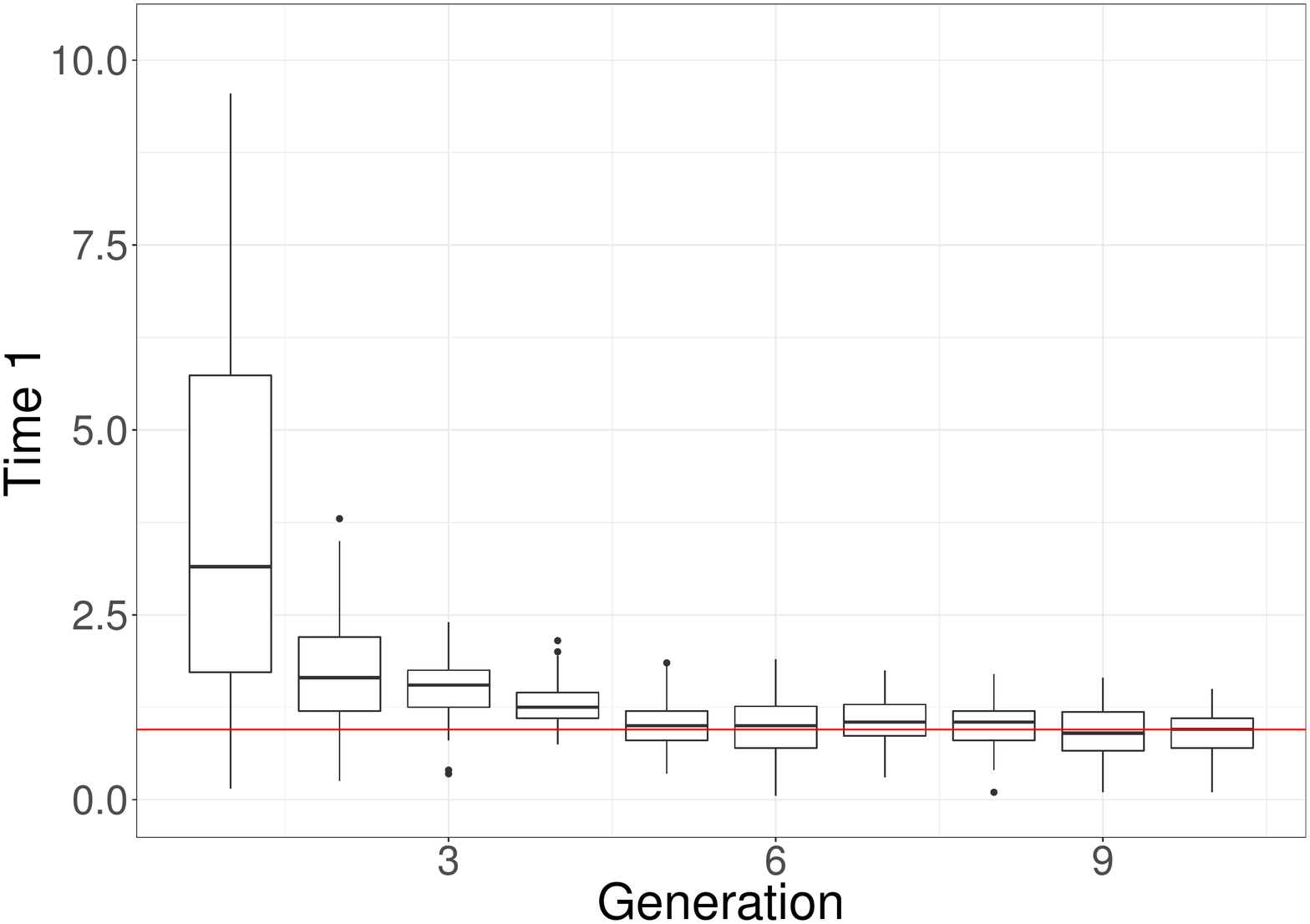}}
		\subfloat[Second observation time.]{\includegraphics[width=0.45\linewidth]{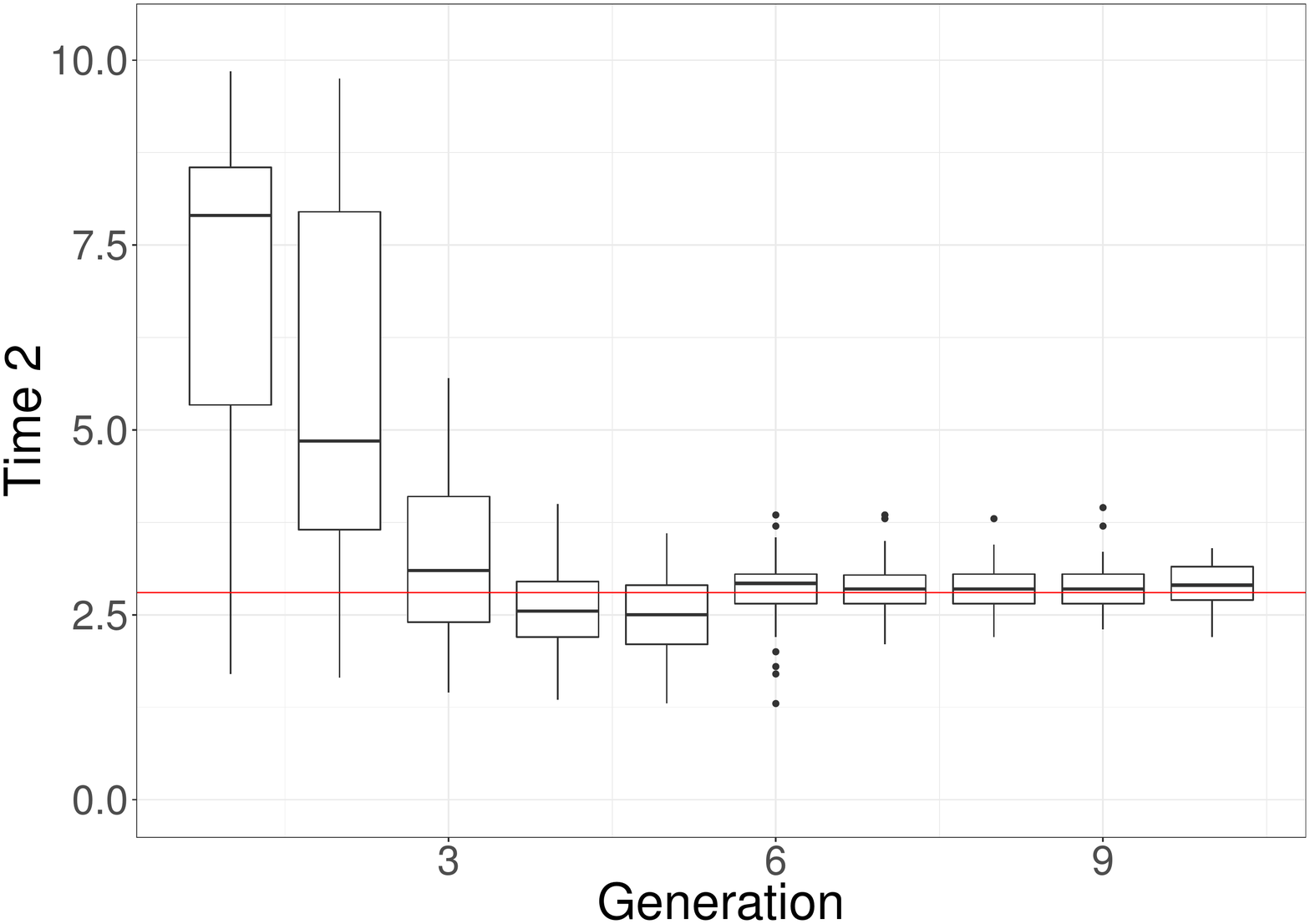}}\\
		\subfloat[Utility.]{\includegraphics[width=0.45\linewidth]{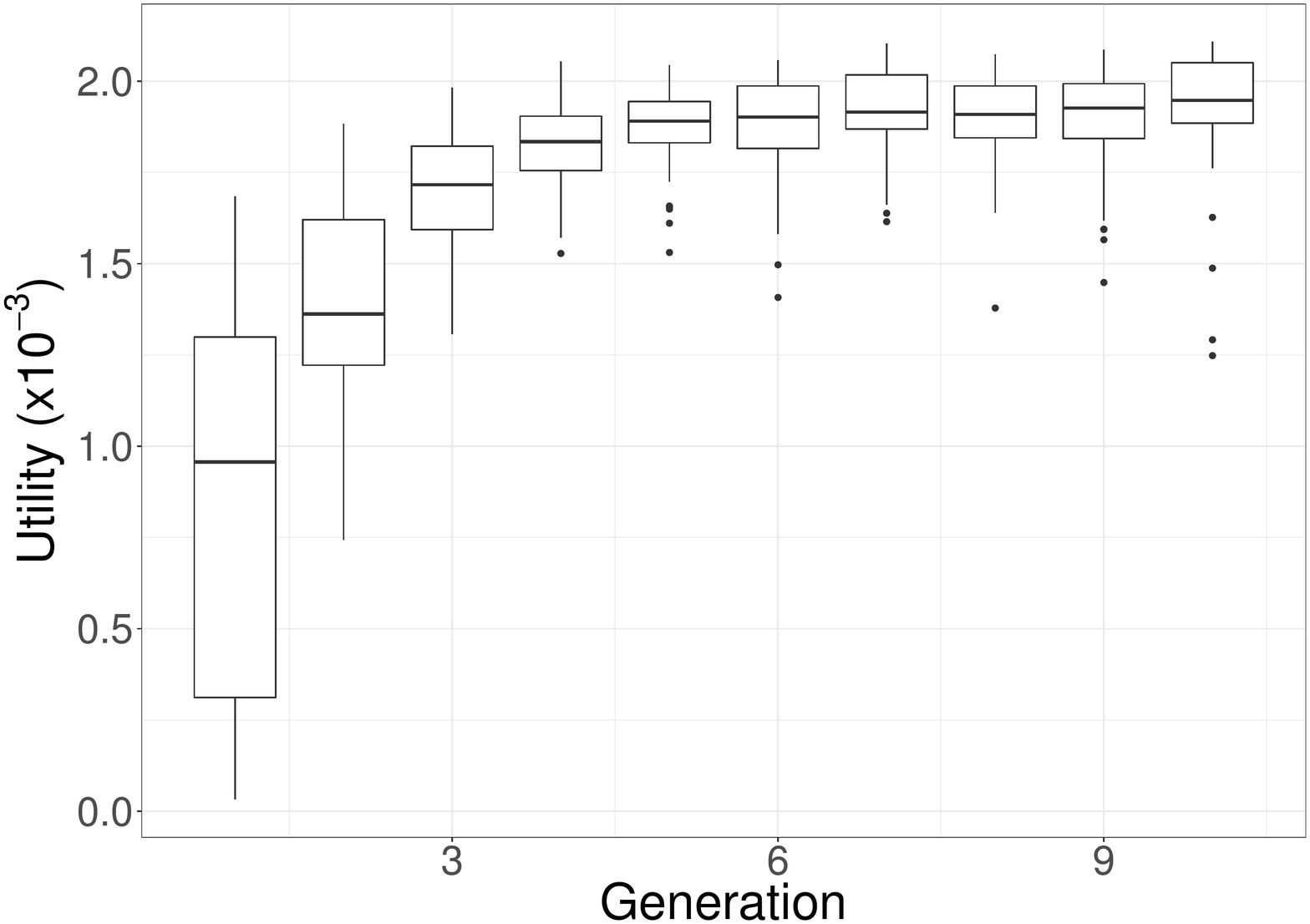}}
	\end{center}
	\caption{Boxplots of the two observation times, and the utility corresponding to the considered designs at each generation of the INSH algorithm. The horizontal lines in (a) and (b) correspond to the optimal observation times evaluated using the ABCdE method.} \label{Death_boxplots}
\end{figure}

Figure \ref{Death_boxplots} demonstrates the convergence of the INSH algorithm to the optimal observation times, and the maximum utility, for two observation times.

\begin{table}[H]
\caption{Comparison of the optimal observation times for the death process, from \citet{Cook:2008}, \citet{Drovandi:2013}, \citet{Price:2016}, and the INSH algorithm. $|t|$ is the pre-determined number of observation times, and $i$ is the  $i^{th}$ time.} \label{table:deathmodel_oeds}
	\begin{center}
		\begin{tabular}{cccccc}
		\hline
			& & \multicolumn{4}{c}{Design Method}\\
			$|t|$ & $i$ & Cook, \emph{et al.}\ & Drovandi $\&$ Pettitt & ABCdE & INSH\\
			\hline
			1 & 1 & 1.70 & 1.60 & 1.50 & 1.45 \\
			\hline
			2 & 1 & 0.90 & 1.15 & 0.80 & 0.95 \\
			- & 2 & 2.40 & 3.05 & 2.80 &  2.80 \\
			\hline
			3 & 1 & 0.70 & 0.75 & 0.40 & 0.60 \\ 
			- & 2 & 1.50 & 1.90 & 1.30 &  1.15 \\
			- & 3 & 2.90 & 3.90 & 2.60 &  2.70 \\
			\hline
		 	4 & 1 & 0.80 & 0.75 &  0.30 & 0.10 \\
			-  & 2 & 1.70 & 1.70 & 0.70 &  0.50 \\
			-  & 3 & 3.10 & 2.75 & 1.30 &  1.20 \\
			-  & 4 & 5.30 & 4.35 & 2.70 & 2.75 \\
			\hline
			6 & (1,2) & - & - &  - & (0.05,0.15) \\
			- & (3,4) & - & - &  - & (0.45,1.15) \\
			- & (5,6) & - & - &  - & (1.75,3.05) \\
			\hline
			8 & (1,2) & - & - &  - & (0.05,0.15) \\
			- & (3,4) & - & - &  - & (0.25,0.45) \\
			- & (5,6) & - & - &  - & (0.80,1.40) \\
			- & (7,8) & - & - &  - & (2.20,2.90) \\
			\hline
		\end{tabular}
	\end{center}
\end{table}
Table \ref{table:deathmodel_oeds} contains the optimal experimental designs for different numbers of observations of the Markovian death model, evaluated by \citet{Cook:2008}, \citet{Drovandi:2013}, \citet{Price:2016} (where computationally feasible), and the INSH algorithm.

Table \ref{death:insh_pars} contains the input parameters for the INSH algorithm, applied to the death model.

\begin{table}[H]
\caption{Input parameters for the INSH algorithm, applied to the Markovian death model. Note that $m_w$ and $r_w$ are applied each for $W/2$ iterations.} \label{death:insh_pars}
	\begin{center}
		\begin{tabular}{rrrrr}
		$|t|$ & $W$ & $m_w$ & $r_w$ & No. initial designs \\
		\hline
		1 & 8 & $(3, 5)$ & $(10,6)$ & 20 \\
		2 & 10 & $(3, 5)$ & $(20,12)$ & 50 \\
		3 & 16 & $(3, 5)$ & $(20,12)$ & 120 \\
		4 & 20 & $(3, 5)$ & $(20,12)$ & 250 \\
		6 & 30 & $(3, 5)$ & $(25,15)$ & 400 \\
		8 & 50 & $(3, 5)$ & $(25,15)$ & 600 \\
		\end{tabular}
	\end{center}
\end{table}

\newpage
\section*{Appendix C}
\textit{INSH Results for the Pharmacokinetic Model}

\begin{figure}[H]
	\begin{center}
		\includegraphics[width=0.8\linewidth]{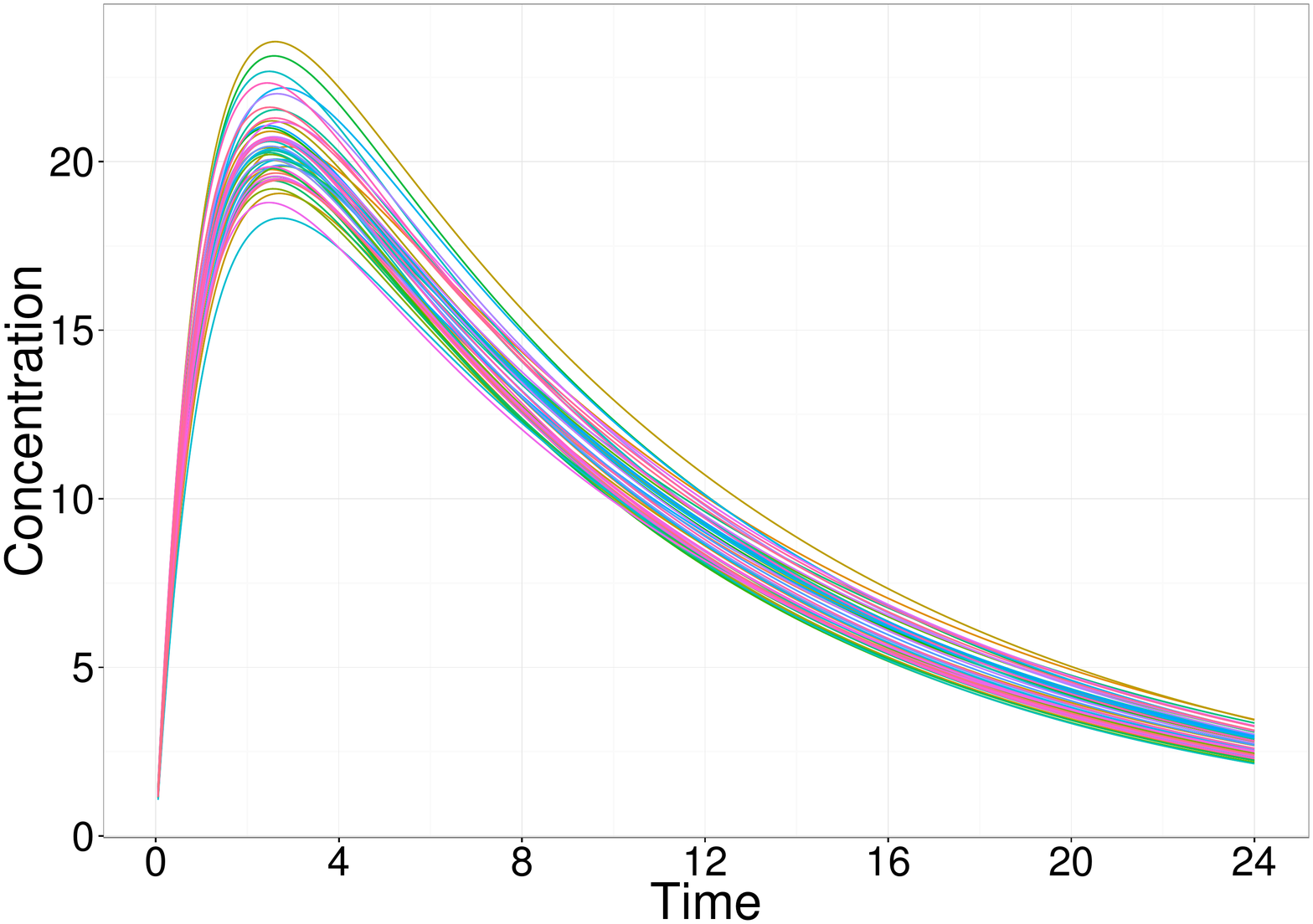}
	\end{center}
	\caption{Plot of 50 mean concentrations over time of the pharmacokinetic model simulated using values sampled from the prior distribution.} \label{PK_50_mean_concentrations}
\end{figure}

Figure \ref{PK_50_mean_concentrations} demonstrates the mean concentrations over time of the pharmacokinetic model evaluated for 50 parameter sets sampled from the prior distribution. 

%
%

Table \ref{PKexample:opttimes} gives the 15 optimal observation times from the top three designs considered by the INSH algorithm. Each chosen optimal design shows the same pattern -- four early observation times ($<1.2$), followed by a cluster of observation times around 4-7, and the remaining observations grouped together towards the final permitted time.

\begin{table}[H]
\caption{Three best sampling schedules evaluated from the INSH algorithm for the pharmacokinetic model.}\label{PKexample:opttimes}
\begin{center}
	\begin{tabular}{lc}
		Original rank &	Design   \\
		\hline
		19 & $(0.1961, 0.4840, 0.7506, 1.176, 4.069, 4.780, 5.281,$   \\
		 & $ 6.030, 6.377, 18.22, 18.85, 19.72, 20.33, 21.52, 22.04)$  \\
		 \\
		 2 & $(0.2460, 0.5054, 0.8017, 1.211, 4.035, 4.477, 5.173, 6.101, $   \\
		  & $ 6.632, 17.82, 18.63, 19.71, 20.32, 21.57, 21.98)$ \\
		 \\
		 3 &$(0.1989, 0.4801, 0.7778, 1.103, 4.465, 4.754, 5.776, 6.270,$  \\
		  &$  6.754, 18.50, 18.99, 20.19, 20.87, 21.16, 21.87)$  \\
		\end{tabular}
	\end{center}
\end{table}

\begin{figure}[H] 
	\begin{center}
		\includegraphics[width=0.99\linewidth]{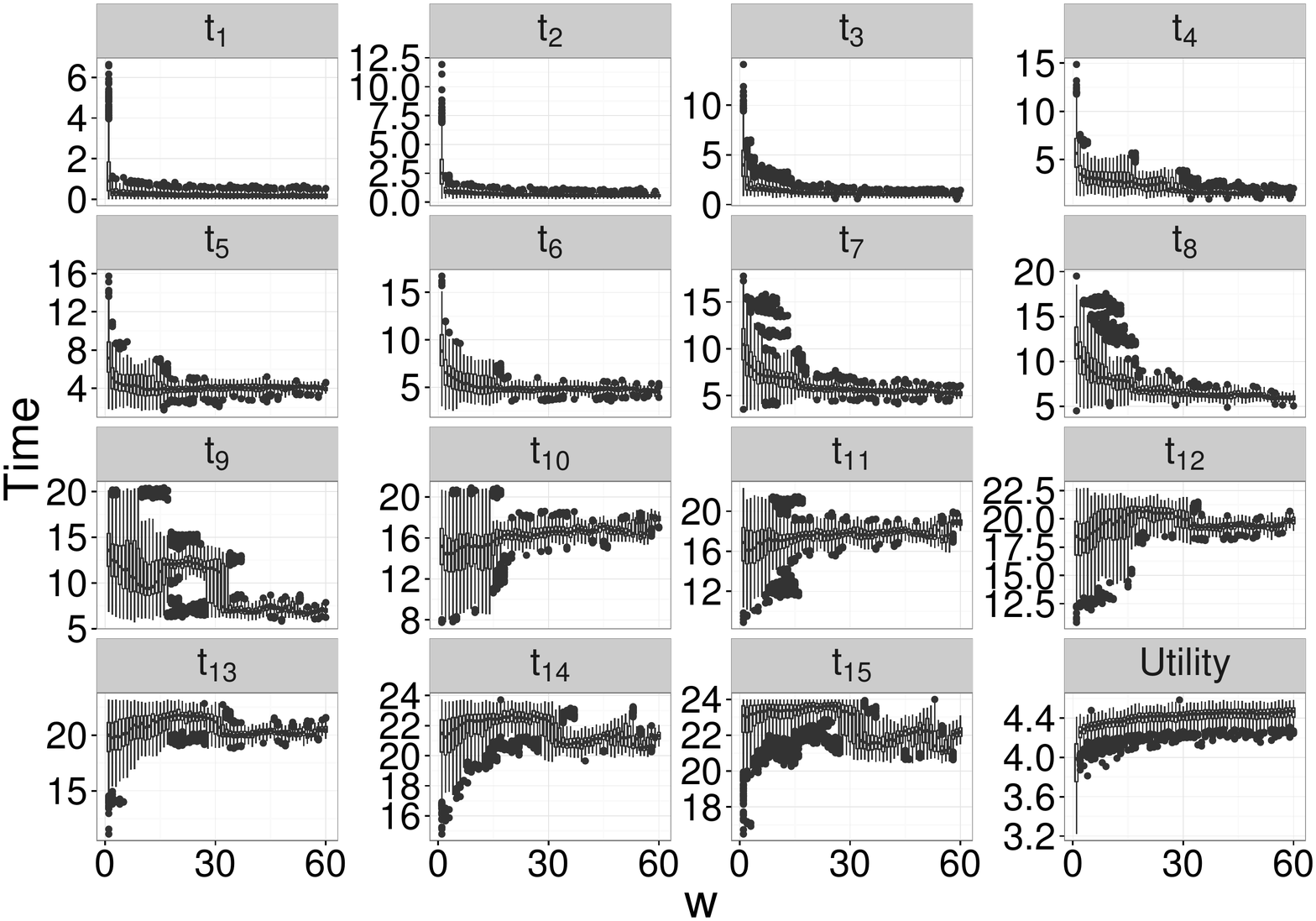}
		\caption{Figure showing the convergence of the sampled designs towards the region near the optimal design. Each panel represents an individual aspect of the sampled designs, the x-axis is the iteration of the INSH algorithm, and the y-axis is the value of the design aspect. The final panel shows the utilities corresponding to the sampled designs.} \label{INSH_pk_convergence}
	\end{center}
\end{figure}

Figure \ref{INSH_pk_convergence} shows box plots of the observation times, and the utility evaluations of the corresponding designs considered at each wave of the INSH algorithm. The figure for $t_9$, for example, depicts the ability of the INSH algorithm to search multiple regions simultaneously. In particular, iterations 17-27 are considering observation times in approximately three clusters -- around times of 5, 12 and 15.

Table \ref{table_pk_klds} contains the estimated expected utility, median utility, and the $10^{th}$ and $90^{th}$ percentiles, corresponding to each of the top 20 designs considered by INSH, and the optimal returned by the ACE algorithm.

\begin{table}[H]
\caption{Summary statistics of estimated utilities corresponding to the top 20 designs from the INSH algorithm, and the optimal design returned by the ACE algorithm. The design highlighted in bold is the design considered to be the optimal from the INSH algorithm.} \label{table_pk_klds}
\centering
\begin{tabular}{r|cccc}
  \hline
  & \multicolumn{4}{c}{Utility}\\
 Design & Mean & Median & 10\% & 90\% \\ 
  \hline
  ACE & 4.4987 & 4.5004 & 4.4844 & 4.5102 \\
1 & 4.4874 & 4.4865 & 4.4715 & 4.5040 \\ 
  2 & 4.4725 & 4.4710 & 4.4596 & 4.4894 \\ 
  3 & 4.4707 & 4.4685 & 4.4598 & 4.4864 \\ 
  4 & 4.4700 & 4.4686 & 4.4576 & 4.4835 \\ 
  5 & 4.4991 & 4.4995 & 4.4870 & 4.5111 \\ 
  6 & 4.4739 & 4.4719 & 4.4638 & 4.4900 \\ 
  7 & 4.4707 & 4.4733 & 4.4558 & 4.4866 \\ 
  8 & 4.5034 & 4.5015 & 4.4956 & 4.5156 \\ 
  9 & 4.4633 & 4.4648 & 4.4506 & 4.4758 \\ 
  10 & 4.4595 & 4.4633 & 4.4444 & 4.4736 \\ 
  11 & 4.4652 & 4.4633 & 4.4526 & 4.4803 \\ 
  12 & 4.4733 & 4.4742 & 4.4608 & 4.4868 \\ 
  13 & 4.4508 & 4.4497 & 4.4349 & 4.4654 \\ 
  14 & 4.4748 & 4.4754 & 4.4616 & 4.4878 \\ 
  15 & 4.4702 & 4.4690 & 4.4527 & 4.4941 \\ 
  16 & 4.4537 & 4.4523 & 4.4426 & 4.4725 \\ 
  17 & 4.4625 & 4.4633 & 4.4439 & 4.4846 \\ 
  18 & 4.4853 & 4.4877 & 4.4702 & 4.4991 \\ 
  {\bf 19} & {\bf 4.5052} & {\bf 4.5076} & {\bf 4.4866} & {\bf 4.5204} \\ 
  20 & 4.4799 & 4.4780 & 4.4676 & 4.4975 \\ 
   \hline
\end{tabular}
\end{table}

\begin{figure}[H]
	\begin{center}
		\includegraphics[width=0.99\linewidth]{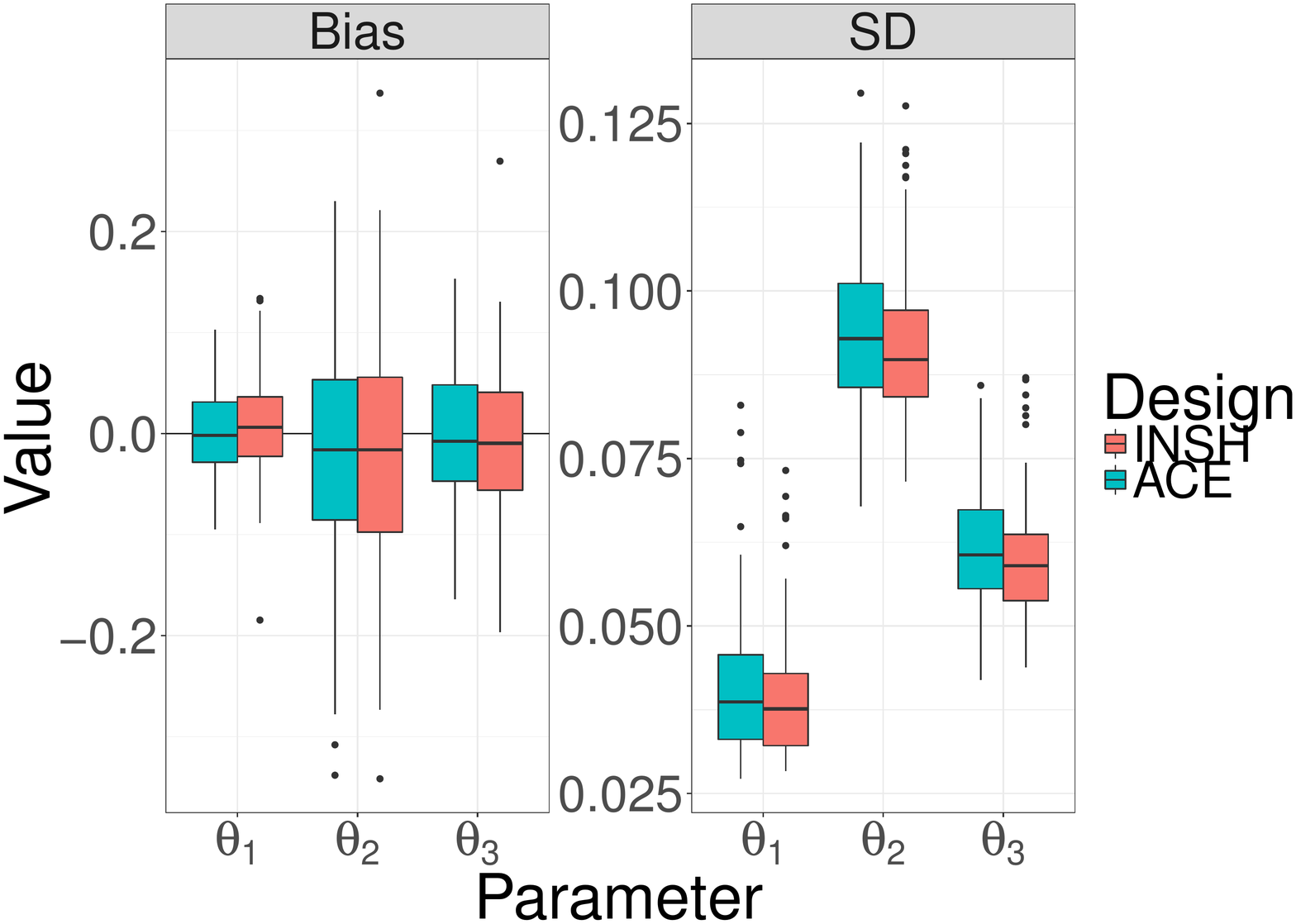}
	\end{center}
	\caption{Comparison of the bias in MAP estimate, and posterior standard deviation of each parameter in the pharmacokinetic model.} \label{PK_inference}
\end{figure}

Figure \ref{PK_inference} provides a comparison of the inferential performance of the two optimal designs -- corresponding to INSH and ACE -- with regards to bias in a point estimate, and the posterior standard deviation. We simulated 100 experiments from random parameters drawn from the prior distribution, and evaluated an approximate posterior distribution using a Metropolis-Hastings algorithm (retaining 100{,}000 samples from the posterior, following a burn-in of 10{,}000). The bias is estimated as the difference between the MAP (\emph{maximum a posteriori}) estimate and the true parameter value that created the simulated data. Recall, the prior variance was 0.05 for each parameter (prior standard deviation is approximately 0.224).

It appears as though the design evaluated by the INSH algorithm performs marginally better with respect to the posterior standard deviation -- that is, the estimated standard deviations are slightly lower for each parameter. The bias in the parameter estimates appears roughly equivalent between the two designs.

\newpage
\section{Appendix D}
\textit{INSH Results for the Logistic Regression Example}

Table \ref{LR_INSH_pars} contains the choices of parameters $r_w$ and $m_w$ for the INSH algorithm. Overall, for $n=6,10$ and $24$, $W=120, 132$ and $240$ iterations were used, and $r_w$ and $m_w$ were chosen such that a total of 600 designs were evaluated at each iteration. For $n=48$, $W=360$, and a total of 1200 designs were considered at each iteration, and more emphasis was placed on exploration early on -- to account for the larger-dimensional design space. In each example, we initiated the INSH algorithm with 10000 designs -- with probability 0.5, uniformly sampled from the design space, otherwise, on a boundary (i.e., all elements of the design consisted of randomly selected -1's and 1's).

\begin{table}[H]
\caption{INSH algorithm parameter choices for the Logistic Regression example.}
\label{LR_INSH_pars}
	\begin{center}
		\begin{tabular}{cccc}
		$n$ & Parameter & Values & Iterations per value\\
		\hline
			& $r_w$ & (200,100,50 ,25,15,10) &  \\
	  $6$ & $m_w$ & $(3, 6,12,24,40,60)$ & 20 \\ 
			& $\sigma_w$ & (0.20, 0.10, 0.05, 0.025, 0.01, 0.005) & \\
			\hline
			& $r_w$ & (200,100,50 ,25,15,10) &  \\
	  $10$ & $m_w$ & $(3,6,12,24,40, 60)$ & 22 \\ 
			& $\sigma_w$ & (0.20, 0.10, 0.05, 0.025, 0.01, 0.005) & \\
			\hline
			& $r_w$ & (200,100,50 ,25,15,10) &  \\
	  $24$ & $m_w$ & $(3,6,12,24,40, 60)$ & 40 \\ 
			& $\sigma_w$ & (0.20, 0.10, 0.05, 0.025, 0.01, 0.005) & \\
			\hline
			& $r_w$ & (200,100,50 ,25,15,10) &  \\
	  $48$ & $m_w$ & $(6,12,24,48,80,120)$ & 60 \\ 
			& $\sigma_w$ & (0.20, 0.10, 0.05, 0.025 0.010, 0.0025) & \\
		\end{tabular}
	\end{center}
\end{table}

Tables \ref{LR_OD_n6}, \ref{LR_OD_n10}, \ref{LR_OD_n24} and \ref{LR_OD_n48} show the optimal designs found by the INSH algorithm for $n=6,10,24$, and $48$, respectively.
\begin{table}[H]
\caption{Optimal design from the INSH algorithm for the Logistic regression model with $n=6$.} \label{LR_OD_n6}
	\begin{center}
		\begin{tabular}{c|cccc}
			$n$ & $x_1$ & $x_2$ & $x_3$ & $x_4$ \\
			\hline
			1 & -0.80 & 0.98 & 0.99 & 0.98 \\ 
			2 & 1.00 & -0.47 & 0.97 & -0.99 \\ 
			3 & 1.00 & -0.63 & 1.00 & 1.00 \\ 
			4 & -0.89 & 1.00 & 0.56 & -1.00 \\ 
			5 & 0.81 & -1.00 & -0.99 & -0.85 \\ 
			6 & -0.97 & 0.45 & -1.00 & 0.99 \\ 
		\end{tabular}
	\end{center}
\end{table}

\begin{table}[H]
\caption{Optimal design from the INSH algorithm for the Logistic regression model with $n=10$.} \label{LR_OD_n10}
\begin{center}
		\begin{tabular}{c|cccc}
			$n$ & $x_1$ & $x_2$ & $x_3$ & $x_4$ \\
			\hline
			1 & -0.80 & 1.00 & 1.00 & 0.98 \\ 
			2 & 0.72 & -1.00 & -0.98 & 0.99 \\ 
			3 & -1.00 & 0.53 & -0.86 & -0.99 \\ 
			4 & 1.00 & -0.85 & 0.39 & -1.00 \\ 
			5 & 0.97 & -0.32 & 0.99 & -0.92 \\ 
			6 & 0.91 & -0.46 & 1.00 & 0.99 \\ 
			7 & -0.72 & 1.00 & 0.90 & -0.97 \\ 
			8 & -1.00 & 0.51 & -0.93 & 0.99 \\ 
			9 & 0.92 & -0.99 & -1.00 & -0.87 \\ 
			10 & -0.97 & 0.64 & -0.99 & -0.96 \\ 
\end{tabular}
\end{center}
\end{table}

\begin{table}[H]
\caption{Optimal design from the INSH algorithm for the Logistic regression model with $n=24$.} \label{LR_OD_n24}
	\begin{center}
		\begin{tabular}{c|cccc}
			$n$ & $x_1$ & $x_2$ & $x_3$ & $x_4$ \\
			\hline
                          1 & -0.33 & 0.50 & 0.89 & 0.89 \\ 
                          2 & -0.25 & 0.78 & 0.89 & -0.96 \\ 
                          3 & -0.95 & 0.94 & 0.93 & 0.97 \\ 
                          4 & 0.68 & -0.98 & -0.92 & -0.81 \\ 
                          5 & -0.93 & 0.99 & 0.86 & -0.88 \\ 
                          6 & -0.77 & 0.98 & 0.82 & 0.87 \\ 
                          7 & 0.82 & -0.96 & -0.83 & -0.99 \\ 
                          8 & -0.64 & 0.11 & -0.66 & 0.98 \\ 
                          9 & -0.48 & 0.86 & 0.96 & 0.92 \\ 
                          10 & -0.76 & 0.40 & -0.95 & -0.94 \\ 
                          11 & 1.00 & -0.74 & 0.97 & 1.00 \\ 
                          12 & -0.99 & 0.46 & -0.97 & 0.96 \\ 
                          13 & 0.96 & -0.44 & 0.97 & -0.97 \\ 
                          14 & 0.78 & -0.93 & -0.89 & 0.96 \\ 
                          15 & -0.34 & 0.74 & 0.99 & -0.98 \\ 
                          16 & 1.00 & -0.95 & -0.92 & -0.99 \\ 
                          17 & -0.22 & -0.14 & -0.86 & -0.96 \\ 
                          18 & -0.96 & 0.85 & -0.59 & -0.81 \\ 
                          19 & -0.70 & 0.12 & -0.98 & -0.93 \\ 
                          20 & -0.74 & 0.30 & -0.98 & 0.96 \\ 
                          21 & 0.99 & -0.50 & 0.83 & -0.60 \\ 
                          22 & 0.53 & -0.99 & -0.91 & 0.99 \\ 
                          23 & -0.21 & -0.26 & -1.00 & 0.85 \\ 
                          24 & 1.00 & -0.55 & 0.97 & 0.96 \\ 
		\end{tabular}
	\end{center}
\end{table}

\begin{table}[H]
\caption{Optimal design from the INSH algorithm for the Logistic regression model with $n=48$.} \label{LR_OD_n48}
	\begin{center}
		\begin{tabular}{c|cccc}
			$n$ & $x_1$ & $x_2$ & $x_3$ & $x_4$ \\
			\hline
                            1 & -0.98 & 0.99 & 1.00 & -0.60 \\ 
                            2 & -0.70 & 0.67 & 0.73 & 0.99 \\ 
                            3 & -0.97 & 0.95 & 0.38 & -0.95 \\ 
                            4 & -0.95 & 0.92 & -0.18 & 0.94 \\ 
                            5 & 0.76 & -0.50 & 0.94 & -0.88 \\ 
                            6 & 0.95 & -0.95 & -0.60 & 0.98 \\ 
                            7 & -1.00 & 0.38 & -1.00 & -0.88 \\ 
                            8 & 0.42 & -0.84 & -0.94 & 0.89 \\ 
                            9 & -0.91 & 0.75 & -0.85 & -0.98 \\ 
                            10 & 0.51 & -0.95 & -0.99 & 1.00 \\ 
                            11 & 0.52 & -0.13 & 0.97 & -0.89 \\ 
                            12 & 0.46 & -0.71 & -0.87 & 0.92 \\ 
                            13 & -0.70 & 0.94 & 0.84 & 0.92 \\ 
                            14 & -0.95 & 0.59 & -1.00 & -0.98 \\ 
                            15 & -0.36 & 0.90 & 0.99 & -0.91 \\ 
                            16 & 0.99 & -0.65 & 0.90 & 0.93 \\ 
                            17 & 0.81 & -0.50 & 0.65 & 0.74 \\ 
                            18 & -0.52 & 1.00 & 0.92 & 0.91 \\ 
                            19 & 0.29 & -0.79 & -0.68 & 0.72 \\ 
                            20 & 0.98 & -0.96 & -0.93 & -0.93 \\ 
                            21 & 0.96 & -0.74 & 0.93 & 0.81 \\ 
                            22 & 0.93 & -0.74 & -0.68 & -0.90 \\ 
                            23 & 0.63 & -0.69 & -0.96 & -0.94 \\ 
                            24 & 0.99 & -0.89 & 0.92 & 1.00 \\ 
                            25 & -0.01 & -0.21 & -0.94 & 0.65 \\ 
                            26 & 0.98 & -0.85 & -0.55 & -0.91 \\ 
                            27 & -0.85 & 0.98 & 0.98 & -0.98 \\ 
                            28 & -0.64 & 0.64 & -0.40 & 0.98 \\ 
                            29 & 0.94 & -0.78 & 0.85 & -0.43 \\ 
                            30 & 0.82 & 0.01 & 0.97 & -0.94 \\ 
                            31 & -0.98 & 0.42 & -0.97 & -0.91 \\ 
                            32 & 0.38 & -0.89 & -1.00 & -1.00 \\ 
                            33 & 0.99 & -0.61 & 0.58 & 0.97 \\ 
                            34 & -0.96 & 0.31 & -0.95 & 0.99 \\ 
                            35 & 0.32 & -0.73 & -0.99 & -0.06 \\ 
                            36 & -0.39 & -0.31 & -1.00 & 0.75 \\ 
                            37 & 0.97 & -0.15 & 0.78 & -1.00 \\ 
                            38 & -0.99 & 0.31 & -0.96 & 0.96 \\ 
                            39 & -0.51 & 0.94 & 0.92 & -1.00 \\ 
                            40 & 0.74 & -0.98 & -0.95 & -0.25 \\ 
                            41 & -0.81 & 0.46 & -0.62 & 0.99 \\ 
                            42 & -0.87 & 0.99 & 0.16 & -0.50 \\ 
                            43 & 0.87 & -0.95 & -0.68 & -0.93 \\ 
                            44 & 0.37 & -0.90 & -0.92 & -0.75 \\ 
                            45 & 0.82 & -0.43 & 0.93 & -0.95 \\ 
                            46 & 0.90 & -0.27 & 1.00 & 0.94 \\ 
                            47 & 0.91 & -0.25 & 0.98 & -0.71 \\ 
                            48 & -0.93 & 0.94 & 0.88 & 0.93 \\ 
		\end{tabular}
	\end{center}
\end{table}

Note that each optimal design contains many values that are close to the boundary values (-1 and 1). The optimal designs reported in \citet{Overstall:2017} can be found in the \verb+acebayes+ package in R, using the command \verb+optdeslrsig(n)+, where $n$ is the number of replicates.

\begin{figure}[H]
	\begin{center}
		\includegraphics[width=0.99\linewidth]{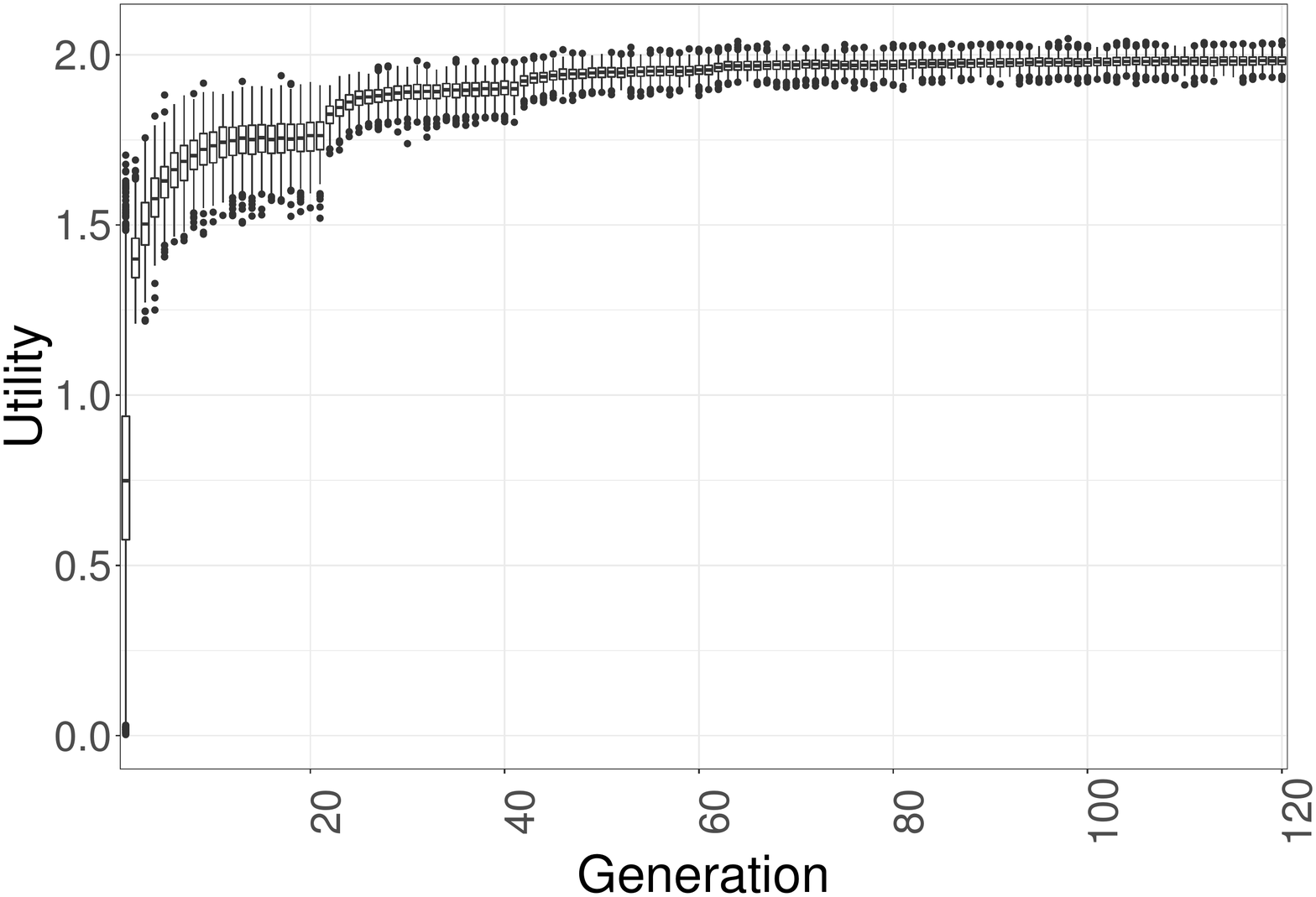}
	\end{center}
	\caption{Box plots of utility evaluations from the INSH algorithm, for the Logistic regression example with $n=6$.} \label{LR_trace_n_6}
\end{figure}

\begin{figure}[H]
	\begin{center}
		\includegraphics[width=0.99\linewidth]{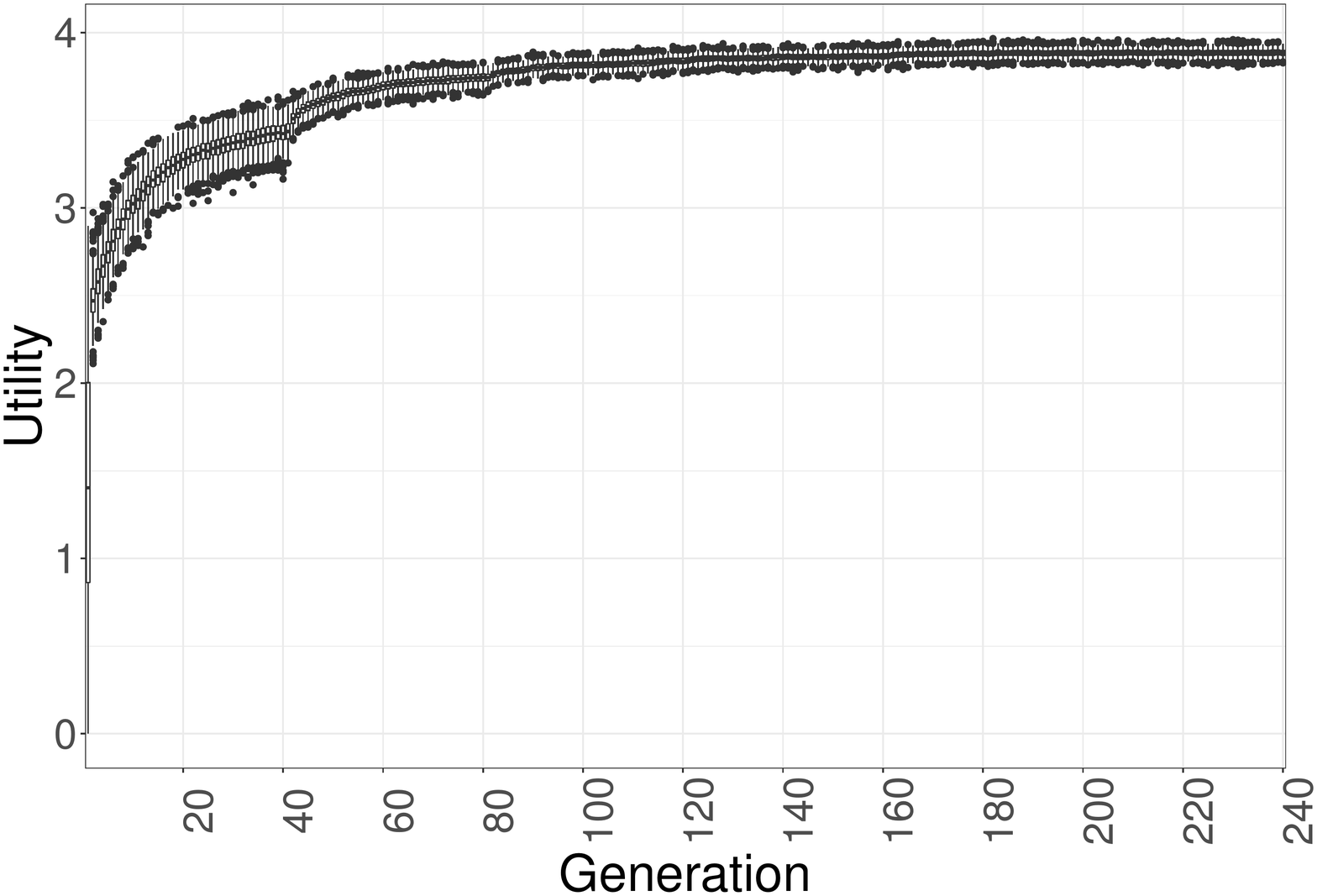}
	\end{center}
	\caption{Box plots of utility evaluations from the INSH algorithm, for the Logistic regression example with $n=24$.} \label{LR_trace_n_24}
\end{figure}

\begin{figure}[H]
	\begin{center}
		\includegraphics[width=0.99\linewidth]{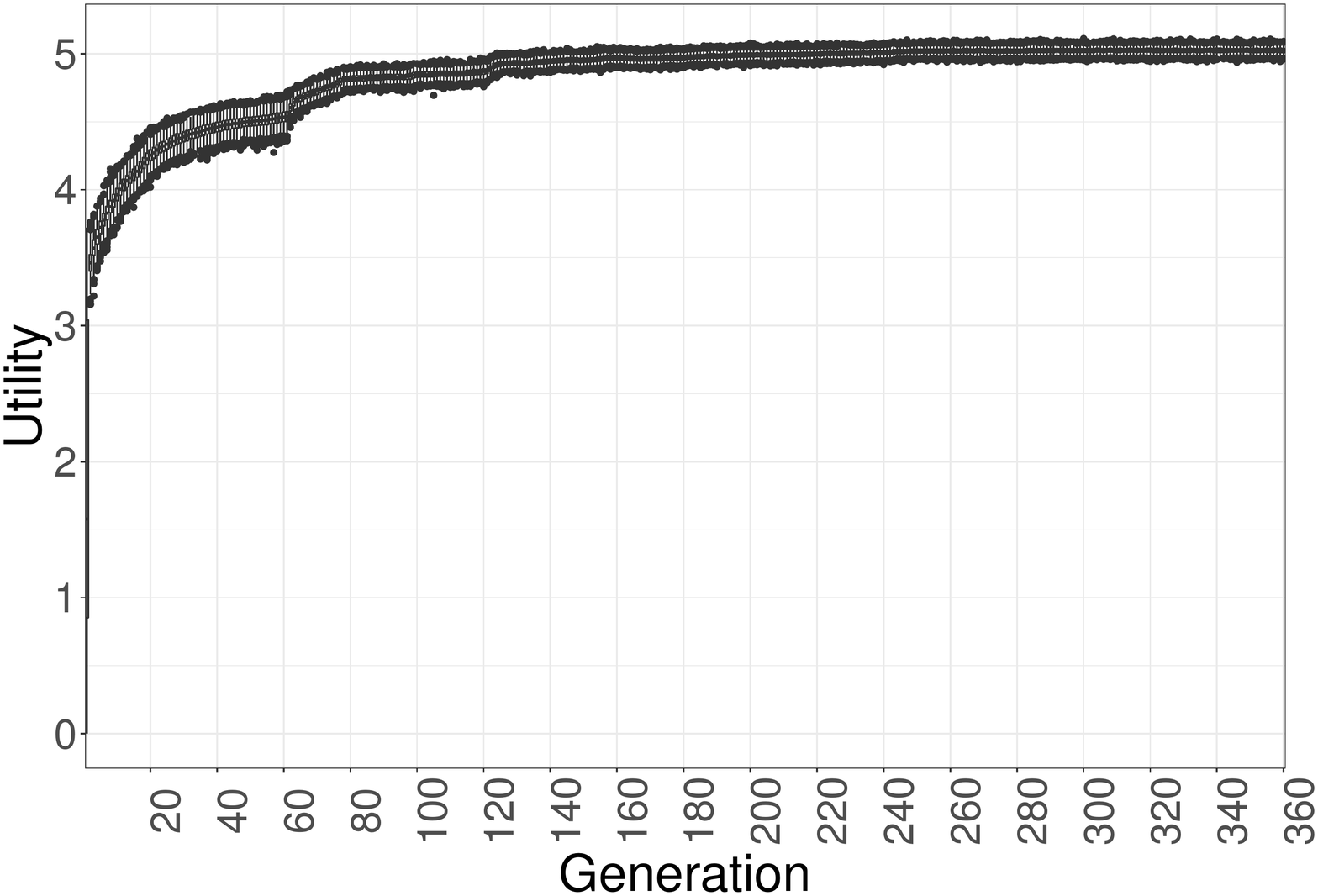}
	\end{center}
	\caption{Box plots of utility evaluations from the INSH algorithm, for the Logistic regression example with $n=48$.} \label{LR_trace_n_48}
\end{figure}

Figures \ref{LR_trace_n_6}, \ref{LR_trace_n_24}, and \ref{LR_trace_n_48} show the progression of the INSH algorithm for the Logistic regression example with $n=6,24$, and $48$, respectively. It appears as though the algorithm has converged to a optimal design region in each case.

\begin{figure}[H]
	\begin{center}
		\includegraphics[width=0.99\linewidth]{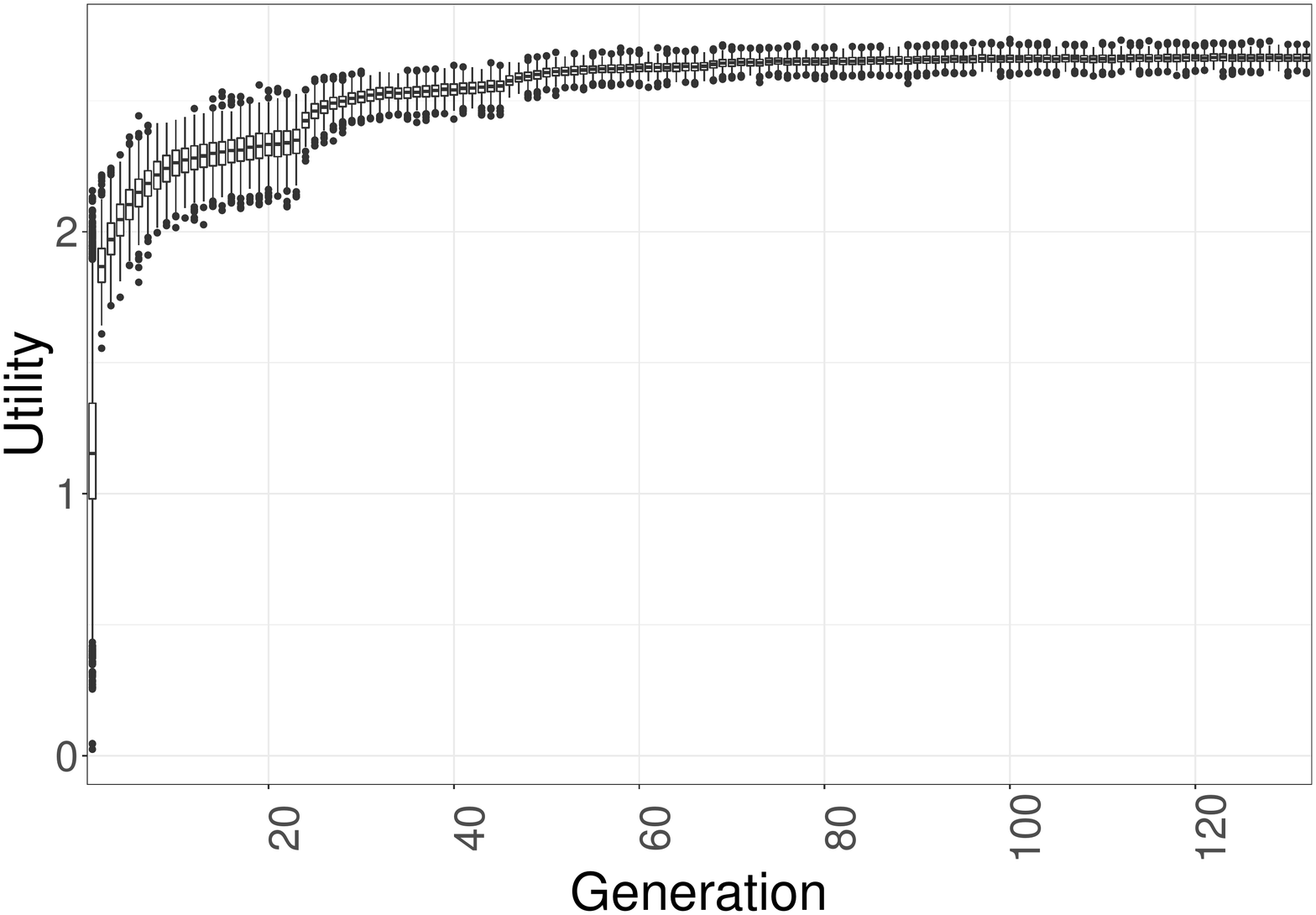}
	\end{center}
	\caption{Box plots of utility evaluations from the INSH algorithm, for the Logistic regression example with $n=10$.} \label{LR_trace_n_10}
\end{figure}

\begin{table}[H]
\caption{The average utility (and 2.5, 97.5-percentiles) of the optimal design found via the ACE, INSH, ACE$_N$ and ACE$_B$ algorithms. For each design, 20 evaluations of the utility were made with $\tilde{B}=B=20000$}
\label{LR_ace_insh_opt_utils}
	\begin{center}
		\begin{tabular}{c|cccc}
			 $n$ & ACE & INSH & ACE$_N$ & ACE$_B$ \\ 
			  \hline
			   6 & 1.99 (1.97, 2.00) & 1.99 (1.97, 2.01) & 1.94 (1.92, 1.96) & 1.96 (1.95, 1.97) \\ 
			 10 & 2.67 (2.66, 2.69) & 2.66 (2.65, 2.68) & 2.64 (2.62, 2,66) & 2.68 (2.66, 2.69) \\ 
			 24 & 3.97 (3.95, 3.98) & 3.88 (3.86, 3.90) & 3.88 (3.86, 3.90) & 3.96 (3.94, 3.98) \\ 
			 48 & 5.11 (5.10, 5.12) & 5.01 (4.98, 5.02) & 5.01 (4.98, 5.02) & 5.06 (5.04, 5.09)
		\end{tabular}
	\end{center}
\end{table}

\begin{table}[H]
\caption{Inputs used to run ACE for the same computation time as INSH: Number of phases of the ACE algorithm ($N_I, N_{II}$), and effort used to evaluate the utility in change step ($B_1$) and for fitting the Gaussian process ($B_2$). Numbers are presented as $(N_I,N_{II}),(B_1,B_2)$. The default settings specified by the authors are $(20,100)$, $(20000,1000)$.}
\label{}
	\begin{center}
		\begin{tabular}{c|cc}
		$n$ & ACE$_N$ & ACE$_B$\\
		\hline
		6 & $(4,20),(20000,1000)$ & $(20,100),(10000,500)$ \\
		10 & $(4,20),(20000,1000)$ & $(20,100),(8000,400)$ \\
		24 & $(3,15),(20000,1000)$ & $(20,100),(5000,250)$ \\
		48 & $(5,25),(20000,1000)$ & $(20,100),(10000,500)$ \\
		\end{tabular}
	\end{center}
\end{table}

\end{document}